\documentclass[fleqn,usenatbib]{mnras}
\usepackage{newtxtext,newtxmath}
\usepackage[T1]{fontenc}
\DeclareRobustCommand{\VAN}[3]{#2}
\let\VANthebibliography\thebibliography
\def\thebibliography{\DeclareRobustCommand{\VAN}[3]{##3}\VANthebibliography}
\usepackage{graphicx}
\usepackage{amsmath}
\usepackage{stackengine}
\usepackage{float}
\usepackage{longtable}
\usepackage{cancel}
\usepackage{comment}

\newcommand*\mean[1]{\overline{#1}}

\title[Rotation and activity in HARPS data with multi-GPs]{Modelling stellar variability in archival HARPS data: I - Rotation and activity properties with multi-dimensional Gaussian Processes}

\author[H. Yu et al.]{
Haochuan Yu$^{1}$\thanks{\href{mailto:haochuan.yu@physics.ox.ac.uk}{haochuan.yu@physics.ox.ac.uk}},
Suzanne Aigrain$^{1}$,
Baptiste Klein$^{1}$,
Oscar Barragán$^{1}$,
Annelies Mortier$^{2}$,
\newauthor
Niamh K. O'Sullivan$^{1}$,
and Michael Cretignier$^{1}$
\\
$^{1}$Sub-department of Astrophysics, Department of Physics, University of Oxford, Oxford, OX1 3RH, UK \\
$^{2}$School of Physics \& Astronomy, University of Birmingham, Edgbaston, Birmingham B15 2TT, UK
}

\date{Accepted XXX. Received YYY; in original form ZZZ}
\pubyear{2024}

\begin{document}
\label{firstpage}
\pagerange{\pageref{firstpage}--\pageref{lastpage}}
\maketitle

\begin{abstract}
Although instruments for measuring the radial velocities (RVs) of stars now routinely reach sub-meter per second accuracy, the detection of low-mass planets is still very challenging. The rotational modulation and evolution of spots and/or faculae can induce variations in the RVs at the level of a few m/s in Sun-like stars. To overcome this, a multi-dimensional Gaussian Process framework has been developed to model the stellar activity signal using spectroscopic activity indicators together with the RVs. A recently published computationally efficient implementation of this framework, \texttt{S+LEAF~2}, enables the rapid analysis of large samples of targets with sizeable data sets. In this work, we apply this framework to HARPS observations of 268 well-observed targets with precisely determined stellar parameters. Our long-term goal is to quantify the effectiveness of this framework to model and mitigate activity signals for stars of different spectral types and activity levels. In this first paper in the series, we initially focus on the activity indicators (S-index and Bisector Inverse Slope), and use them to a) measure rotation periods for 49 slow rotators in our sample, b) explore the impact of these results on the spin-down of middle-aged late F, G \& K stars, and c) explore indirectly how the spot to facular ratio varies across our sample. Our results should provide valuable clues for planning future RV planet surveys such as the Terra Hunting Experiment or the PLATO ground-based follow-up observations program, and help fine-tune current stellar structure and evolution models.

\end{abstract}




\section{Introduction}

Since the groundbreaking detection of the first exoplanet orbiting a sun-like star, 51 Peg b \citep{1995Natur.378..355M}, using the radial velocity (RV) method, numerous techniques have been developed and campaigns initiated to discover exoplanets with the ultimate goal of finding another habitable world. The RV method continues to be one of the most promising techniques for detecting Earth-like planets, thanks to the development of optical ultra-stable high-resolution échelle spectrographs such as High Accuracy Radial Velocity Planet Searcher \citep[HARPS;][]{2003Msngr.114...20M}, HARPS-N \citep{2012SPIE.8446E..1VC}, ESPRESSO \citep{2021A&A...645A..96P}, EXPRES \citep{2016SPIE.9908E..6TJ}, NEID \citep{2018SPIE10702E..71S} and the upcoming KPF \citep{2016SPIE.9908E..70G} and HARPS-3 \citep{2016SPIE.9908E..6FT}. These new instruments can achieve extreme precision down to 0.3 $\rm{m s^{-1}}$ \citep[e.g.,][]{2020A&A...639A..77S,Faria2022} and even beyond.

However, despite the sub-meter-per-second accuracy achieved by RV instruments, the detection of Earth analogues remains a significant challenge, primarily attributed to stellar activity. For example, in our own solar system, the amplitude of Earth's RV signal is a mere 0.1 m/s, yet RV fluctuations caused by solar activity can exceed several m/s \citep[e.g.,][]{Meunier2010,Haywood2022}. As a result, signals from small, moderately distant planets, similar to Earth, could easily be obscured by the activity of their host star. Accurate modelling of these activity signals is crucial for the successful detection of such planets.

Variations in the RV time-series can arise from a variety of stellar activity processes, often manifesting across different timescales. Predominantly, these fluctuations are the consequence of the magnetic field of the star interplay with the convection. In the following, we discuss several mechanisms believed to be most crucial in inducing variations in the RVs.

On short timescales, the surface of an FGKM star consists of cells where local convection occurs, a phenomenon known as granulation. In each cell, material heated up rises to the surface, causing a blueshift in the lines, while cooler material sinks, causing a redshift. Since the hotter material is generally brighter than the cooler material, averaging the line profiles over the stellar surface produces an asymmetric line profile, resulting in a net blueshift when measuring the line centers. As these cells, or granules, evolve stochastically on the stellar surface, this effect can induce RV variations at the m/s level, with timescales ranging from a few minutes to several hours \citep[e.g.,][]{Dumusque11,2015A&A...583A.118M,2019A&A...625L...6M,2013ApJ...763...95C,2018ApJ...866...55C,2019ApJ...879...55C}. On even shorter timescales, surface granulation and magnetic events can excite oscillations of the star at characteristic frequencies (acoustic modes), known as pressure-mode (p-mode) oscillations \citep[e.g.,][]{1995A&A...293...87K,2008ApJ...687.1180A}. For sun-like stars, p-mode oscillations can induce RV variations with a period of several minutes and an amplitude of around 0.1 to 1 m/s \citep{2018A&A...612A..44S,2019Geosc...9..114C}. Given that the timescales of these effects are normally short compared to the orbital period of the planet of interest, such effects can be filtered out by either extending the exposure time \citep[e.g.,][]{2019AJ....157..163C}, or binning the data to a lower time resolution.

On moderate timescales, effects caused by active regions on the stellar surface become significant. These regions are shaped by magnetic fields, which can interfere with the convection on the stellar surface in two ways. First, when the magnetic field suppresses the local heat transport, it results in dark regions known as spots. Second, if the magnetic field is not strong enough to entirely suppress the heat transport, it alters local opacity which produces brighter regions, often referred to as faculae or plages. These active regions can induce RV variations through two primary mechanisms: the photometric effect, marked by a localized flux alteration, and the inhibition of the convective blueshift effect, characterized by a reduced blueshift on the local line profile. Therefore, disk-integrated line profiles exhibit distortion due to these localized changes, manifesting as shifts in the RV. As these active regions rotate with the surface of the star, the induced RV signal is modulated by such rotation, resulting in a signal with quasi-periodic variations. \citet{2010A&A...512A..39M} found that, for the Sun, the amplitude of such variation in the RV is around 0.4 to 1.4 m/s and is dominated by the convective blueshift effect. The challenge lies in the fact that the rotation periods of Sun-like stars can be of the same order of magnitudes as the orbital period of exoplanets. This similarity led to a few early claims being subsequently refuted \citep[e.g.,][]{Rajpaul16}.

Several techniques have been developed to mitigate such activity signals. One straightforward approach is to compare the Lomb-Scargle periodograms of RV and spectroscopic activity indicators to visually distinguish between stellar and planetary periods. This method is only effective when the data is relatively regular-sampled and the stellar signal remains less evolved. 
An alternative approach is to employ Stacked Bayesian General Lomb-Scargle periodograms \citep{Mortier2017}, which provides a way to assess the stability of signals over time. Stellar signals can be identified as they are unstable and incoherent, in contrast to planetary signals which are consistent and stable.
A more advanced approach is to use a multi-dimensional Gaussian Process framework, which models the RV together with the activity indicators \citep[e.g.,][]{A12,V15,B22}. We will delve deeper into this approach in section \ref{sec:model}. Besides, there are techniques that conduct activity mitigation at earlier stages. For instance, some start from cross-correlation function (CCF) \citep[e.g.,][]{2014MNRAS.444.3220D,2021MNRAS.505.1699C,2022MNRAS.515.3975J,2022ApJ...935...75Z,2022AJ....164...49D}, while others focus on the spectrum level mitigation \citep[e.g.,][]{2017ApJ...846...59D,2017arXiv171101318J,2021A&A...653A..43C,2022A&A...659A..68C}. For more details on existing activity mitigation techniques, we refer readers to \citet{2022AJ....163..171Z} for an overview.

On longer timescales of several years, the influence of magnetic activity cycles becomes prominent. Solar-type stars exhibit cycles of magnetic activity spanning several years, which can potentially affect all the above activity signals. Such cycles have been linked to long-term RV fluctuations over the years \citep{2011A&A...535A..55D}. This means the RV signals attributed to these cycles can be easily confounded with the RV signals from long-period planets. \citet{2019A&A...632A..81M} demonstrated that such effects can be substantially reduced by decorrelating the RV from chromospheric emissions. In addition to the long-term baseline variations, magnetic activity cycles can alter the characteristic period of signals induced by active regions over time. This is because the distribution of active regions across latitudes changes throughout the magnetic activity cycle (e.g., the butterfly diagram of the Sun) \citep[e.g.,][]{Foing1988}. If the surface rotation rate varies from equator to pole (differential rotation), the rotation period inferred from the modulation of surface heterogeneities (whether in photometry or spectroscopic indicators) will vary along the magnetic cycle. For a Sun-like differential rotation and butterfly pattern, the active regions move from higher to lower latitudes as the cycle progresses, and their rotation period decreases.

Since 2003, the high-resolution spectrograph HARPS has been observing thousands of stars. The rich archival, high-quality data from HARPS provides a unique opportunity to test and apply state-of-the-art methods for activity modelling and mitigation. In this paper, we apply the multi-dimensional Gaussian Process framework to HARPS observations of 268 well-observed targets. We discuss sample selection, data reduction and pre-processing, as well as the basic properties of the sample in section \ref{sec:sample}. We then briefly introduce the multi-dimensional GP framework and detail its implementation in section \ref{sec:model}. We discuss results regarding stellar rotation in section \ref{sec:rot}, and facular to spot ratio in section \ref{sec:f-to-s}. We summarize our key conclusions and outline future work in section \ref{sec:sum}. 

\section{The HARPS Sample} \label{sec:sample}
\subsection{Sample selection}
We utilized archived data from the HARPS, accessible via the ESO science archive \footnote{\url{http://archive.eso.org/scienceportal/home}}.

We retrieved all publicly available HARPS spectra using the ESO's \texttt{astroquery} \footnote{\url{https://github.com/astropy/astroquery}} \citep{2019AJ....157...98G} package. For every observation and each target, we cross-matched the stellar name and position with their respective counterparts in the \texttt{simbad} database \citep{simbad2000}. This step ensured the removal of potential misclassifications from the ESO archive and close-in binary systems from our data set. Polarimetric data, which only constitute a minor portion of the HARPS archive, have been discarded as they were not reduced automatically by the standard pipeline. We did not implement cuts based on the signal-to-noise ratio (SNR) of the spectra, or cuts based on whether the simultaneous calibration has been applied. We anticipated that these factors would manifest themselves through the uncertainties of the pipeline products, which would be taken into account statistically by our Bayesian framework introduced later. Our preliminary target set included a total of 1438 targets with spectral types spanning from F to M. Their effective temperatures ($T_{\rm{eff}}$) approximately ranged between 2800 K and 7500 K.

\begin{figure}
	\centering
	\includegraphics[width=\columnwidth]{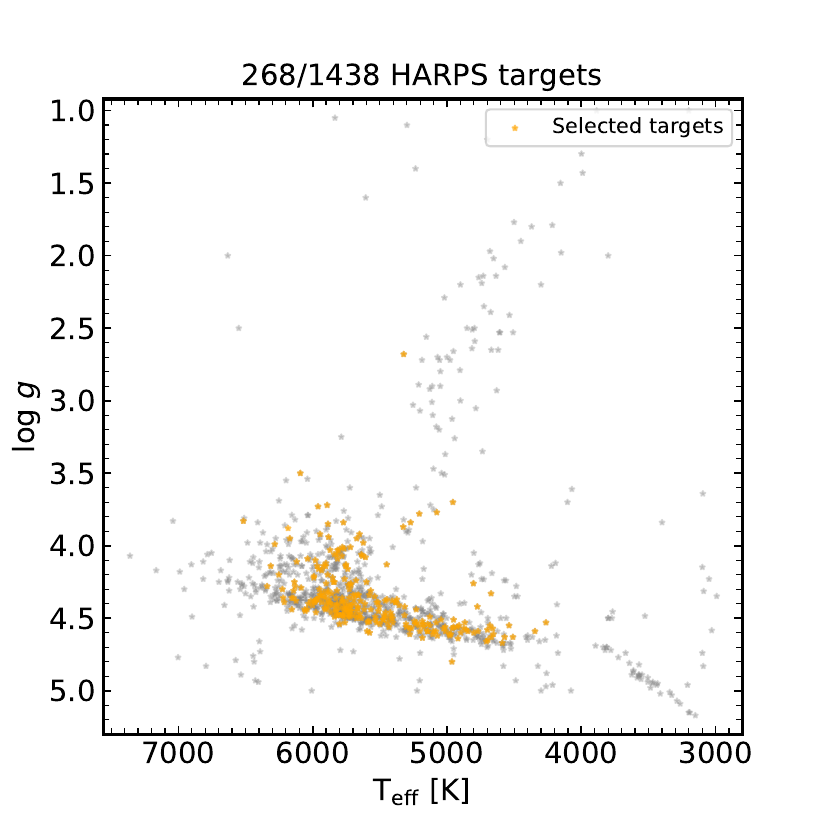}
    \caption{Our sample selection in an HR diagram, with $\rm{log} \ g$ versus $T_{\rm{eff}}$ from GAIA DR2 \citep{2016A&A...595A...1G,2018A&A...616A...1G}. The grey points show the 1438 targets on our initial list. The selected 268 targets in our final sample are highlighted in orange.}
    \label{fig:sample-sel}
\end{figure}

To guarantee a sufficient observation duration for our targets, we excluded those that had less than 40 daily-binned observations. In pursuit of homogeneity in the determination of the stellar parameters, e.g., effective temperature $T_{\rm{eff}}$, surface gravity $\rm{log} \ g$, we restricted our sample to stars listed in the catalogue published in \citet{GdS21}, and used the parameters listed in that catalogue throughout our paper. Our final sample contains 268 targets, with spectral types spanning from F to K. Figure \ref{fig:sample-sel} shows our sample selection in an Hertzsprung-Russell (HR) diagram, with $\rm{log} \ g$ versus $T_{\rm{eff}}$ from GAIA DR2 \citep{2016A&A...595A...1G,2018A&A...616A...1G}. The grey points show the 1438 targets on our initial list, while the selected 268 targets in our final sample are highlighted in orange.

\subsection{Data reduction and pre-processing}
The raw data underwent automatic reduction using version 3.5 and/or 3.8 of the HARPS Data Reduction Software (DRS). For each target, cross-correlation functions (CCFs) were computed by correlating the reduced spectra with a line mask that matches the spectral type of the star. CCF proxies, namely RV and bisector inverse slopes (BIS), were then measured from the CCFs within the DRS. Additionally, we computed a chromospheric emission metric, $S$-index, from the Ca\,{II} H and K lines (resp. 3968.47 and 3933.66~\r{A}) in the S1D spectra, employing the \texttt{ACTIN}\footnote{\url{https://github.com/gomesdasilva/ACTIN}} \citep{GdS18} package.

For each time-series, we calculated the median absolute deviation (MAD) metric, which is considered a more reliable measure of statistical dispersion than the often-employed standard deviation, especially in the presence of outliers. Data points exceeding three times the MAD value were identified as outliers and subsequently excluded. Any time-series extending beyond BJD=2457161.5 was partitioned into two sections for the purpose of outlier removal, given the expectation of two separate baselines resulting from the HARPS fibre upgrade. Finally, we binned each time-series to a maximum of one data point per day.

\subsection{Basic properties of the sample}
\begin{figure*}
    \centering
	\hspace*{-0.2cm}\includegraphics[width=1.0\textwidth]{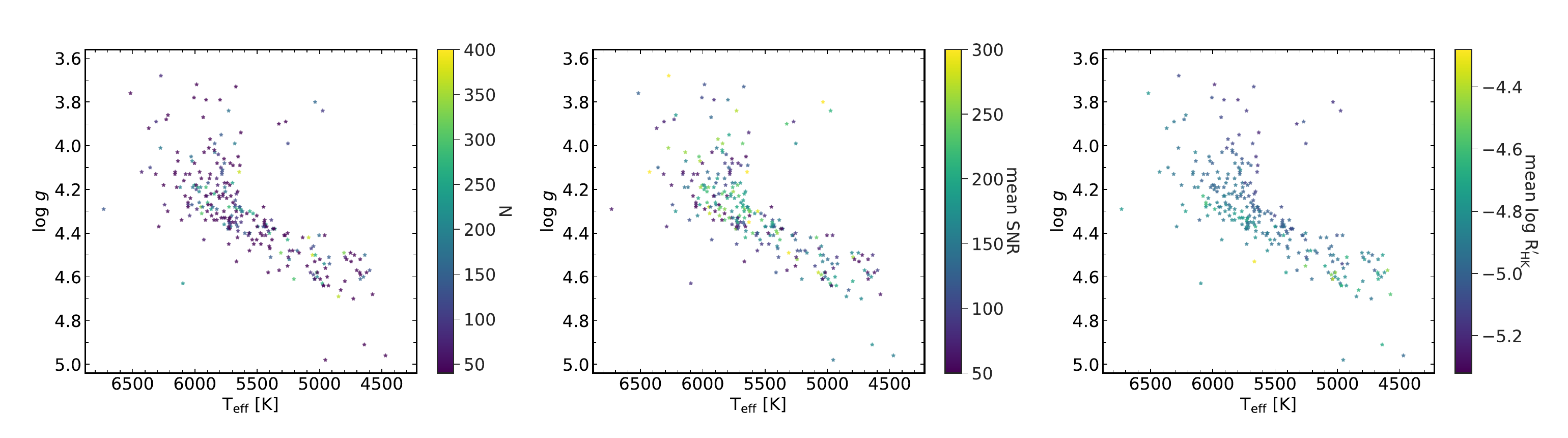}
    \caption{The figure shows the basic properties of the 268 targets in our sample. The targets are displayed in HR diagrams, and the colours of the points in the left, middle, and right panels represent the number of data points in the pre-processed time-series, the mean signal-to-noise ratio (SNR) of the acquired spectra, and the mean $\log R_{\rm{HK}}^{\prime}$ of the star, respectively.}
    \label{fig:sample-prop1}
\end{figure*}

We showcase the basic properties of the 268 targets in our sample using HR diagrams in Figure \ref{fig:sample-prop1}. The colours of the points in the left, middle, and right panels represent the number of data points in the pre-processed time-series, the mean SNR per pixel (in échelle order 50) of the acquired spectra, and the mean $\log R_{\rm{HK}}^{\prime}$ of the star, using values from the catalogue provided by \citet{GdS21}, respectively.
We can see that the numbers of data points in our sample vary, ranging from 40 to over 400, though the majority fall between 40 and 100. The mean SNR spans from 50 to above 300. The mean $\log R_{\rm{HK}}^{\prime}$ for the majority of the targets are below -4.4, suggesting that they are predominantly inactive stars. This aligns with expectations, given that the primary goal of HARPS is exoplanet detection, leading the survey to prioritize low-activity stars.

\begin{figure*}
	\centering
	\includegraphics[width=0.55\textwidth]{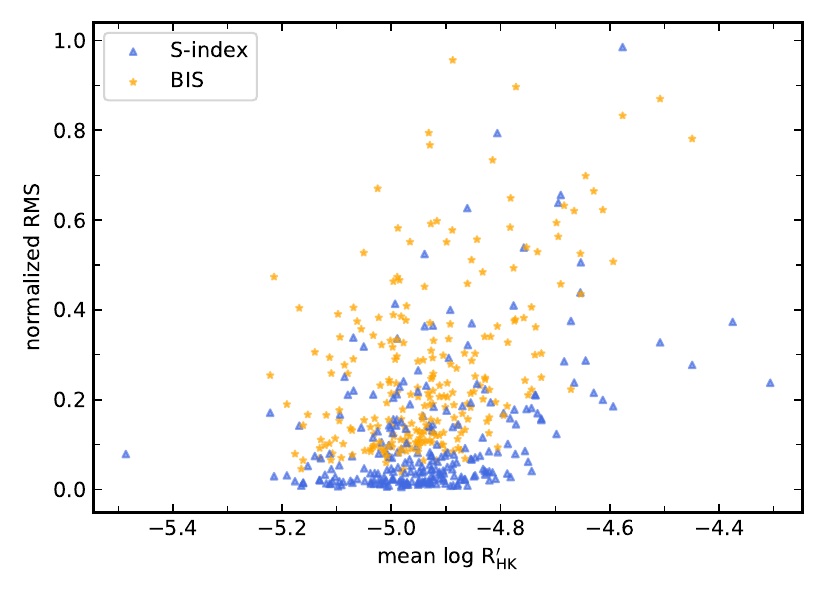}
    \caption{The figure shows the normalized root mean square (RMS) of the time-series for both $S$-index and BIS versus the mean $\log R_{\rm{HK}}^{\prime}$.}
    \label{fig:sample-prop2}
\end{figure*}

In Figure \ref{fig:sample-prop2}, we show the normalized root mean square (RMS) of the time-series for both $S$-index and BIS, plotted against the mean $\log R_{\rm{HK}}^{\prime}$. The RMS values serve as direct metrics of variability in the activity indicators, and as such, they are anticipated to correlate more intimately with the challenges of activity mitigation. We find, however, no clear correlation between the overall RMS values and the mean $\log R_{\rm{HK}}^{\prime}$, even though there appears to be a tentative positive correlation between the RMS's upper envelope and the mean $\log R_{\rm{HK}}^{\prime}$. For instance, at a mean $\log R_{\rm{HK}}^{\prime}$ value of approximately -4.9, the corresponding RMS for either $S$-index or BIS can fluctuate by an order of magnitude. Therefore, we deduce that relying solely on the mean $\log R_{\rm{HK}}^{\prime}$ is inadequate as an activity metric within the context of activity mitigation. We recommend that future survey target selections consider additional metrics alongside mean $\log R_{\rm{HK}}^{\prime}$.

\section{Model for stellar activity} \label{sec:model}
Different processes of stellar activity, e.g., p-mode oscillations, granulation, active regions, and magnetic activity cycles, can induce variations in the time-series of RV and spectroscopic activity indicators at various timescales. In this section, we focus on modelling the variations induced by the active regions — notably the effects of faculae/plages and spots. This is primarily because such activity signal is modulated by the rotation of the star, and the associated timescales (i.e., tens of days) are closest to the orbital periods of the planets. In the following, we introduce the multi-dimensional GP framework to model such activity-induced signals in section \ref{sec:model-GP}, with the aim of measuring the rotation period of the stars. We then detail the implementation of the framework to activity indicators in section \ref{sec:model-implement}.

\subsection{Multi-dimensional Gaussian Processes framework} \label{sec:model-GP}
GPs are commonly used in recent years as a tool to model stellar activity, given their ability to model the data by parameterising its covariance matrix to constrain the characteristics of the data, e.g., period, evolution timescale. This avoids the necessity of knowing the exact deterministic form of the underlying physical processes, which can be extremely hard in the case of modelling stellar activity as it is expected to be stochastic. For comprehensive descriptions of GPs, we refer readers to specialised literature, e.g., \citet{Rasmussen2006}, \citet{Roberts17} and \citet{2023ARA&A..61..329A}.

At its core, a GP model is characterized by a mean function and a kernel function, the latter parameterizing the covariance matrix. Parameters within these kernel functions, termed \textit{hyper-parameters}, shape the characteristics of the modelled data rather than its exact form. For example, a frequent choice for modelling stellar activity is the Quasi-Periodic (QP) kernel,
\begin{equation} \label{eq:QPkernel}
k_{Q P}\left(t, t^{\prime}\right)=A \exp \left[-\Gamma \sin ^{2}\left(\frac{\pi\left(t-t^{\prime}\right)}{P}\right)-\frac{\left(t-t^{\prime}\right)^{2}}{2 l^{2}}\right] \mathrm{,}
\end{equation}
where $A$ is the amplitude representing the overall scale of the variation from the mean function, $P$ is the characteristic period of the variation, $\Gamma$ is the harmonic complexity, which in simpler terms indicates how much the variation strays from a pure sine oscillation within a given period. $l$ is the 'evolution timescale', which scales with the maximum distance of two data points that are strongly correlated.

Aiming to maximise the usage of information from the spectroscopic time-series in modelling the stellar activity, we want to model the activity signals from the RVs together with a set of selected spectroscopic activity indicators, i.e., $S$-index and BIS. Following \citet{A12,V15,B22,D22}, we assume a set of N observable time-series, $Y_{i=1, \ldots, N} (t)$, which can be time-series of RV and activity indicators, and each time-series $Y_{i}$ follows
\begin{equation}
Y_{i}(t)=f_{i}(t) + [a_{i} G(t)+b_{i} \dot{G}(t)] + \epsilon_{i}(t) \mathrm{,}
\end{equation}
where $f_{i}(t)$ is the deterministic part of the model. For RV, it is where the mean function and planet-induced RV are incorporated. For activity indicators, it is simply the mean function. $\epsilon_{i}(t)$ is the measurement 'white' noise, which normally includes photon noise, calibration noise, etc. Every $Y_{i}(t)$ has a shared entity, $G(t)$. We interpret this as a latent GP variable, approximating the proportion of the visible stellar disc blanketed by active regions. Commonly, a GP with a quasi-periodic (QP) kernel is the modelling choice for this variable. $\dot{G}(t)$ is the first temporal derivative of $G(t)$ and remains a GP, roughly representing how the active regions evolve in time on the stellar disc. Coefficients $a_{i}$ and $b_{i}$ are free parameters that harmonise the interplay between the latent GP variables, $G$ and $\dot{G}$, and various observable time-series, $Y_{i}$. 

Building on this framework, a fully parameterised covariance matrix $\mathbf{K}$ for the N time-series $Y_{i=1, \ldots, N} (t)$ can be established. With a known mean function $\mathbf{m}$ and covariance matrix $\mathbf{K}$, the likelihood function can be written in analytic expression as
\begin{equation}
\mathcal{L}=\frac{1}{\sqrt{2 \pi|\mathbf{K}|}} \exp \left(-\frac{1}{2}(\mathbf{y}-\mathbf{m})^{\mathrm{T}} \mathbf{K}^{-1}(\mathbf{y}-\mathbf{m})\right),
\end{equation}
where $\mathbf{y}$ encompasses the vector of N observed time-series $Y_{i}(t)$. Full Bayesian frameworks, such as Monte Carlo Markov Chain (MCMC) or nested sampling, can be implemented with $\mathcal{L}$ to explore the posterior distributions of free parameters in the model. We direct readers to \citet{V15} and \citet{B22} for details.

For $G(t)$, the traditional approach involving a quasi-periodic (QP) kernel, as outlined in Equation \ref{eq:QPkernel}, faces the drawback of computational costs that rise cubically in relation to the size of the dataset. An alternative option is to use a fast approximated kernel, such as a Matern 3/2 exponential periodic (MEP) kernel facilitated in \texttt{S+LEAF~2}\footnote{\url{https://gitlab.unige.ch/Jean-Baptiste.Delisle/spleaf}} \citep{D22}, of which the computational cost only scales linearly. Such efficiency becomes indispensable when tasked with analysing expansive datasets for a large volume of targets without straining computational resources. We refer readers to \citet{D22} for details on the MEP kernel.

\subsection{Implementation} \label{sec:model-implement}
We implemented the multi-dimensional GP framework on every target in our sample. As we initially focused on the activity indicators in the present work, the exact form of the framework we used can be outlined as follows
\begin{equation}
\begin{aligned}
&\mathrm{BIS}(t)=f_{\mathrm{BIS}}(t) + a_{\mathrm{BIS}} G(t)+b_{\mathrm{BIS}} \dot{G}(t) + \epsilon_{\mathrm{BIS}}(t) \mathrm{,}\\
&S\mathrm{-index}(t)=f_{S\mathrm{-index}}(t) + a_{S\mathrm{-index}} G(t) + \epsilon_{S\mathrm{-index}}(t) \mathrm{,}
\end{aligned}
\end{equation}
where
\begin{equation}
\begin{aligned}
&f_{\mathrm{BIS}}(t) = \mean{\mathrm{BIS}(t)}\mathrm{,} \\
&f_{S\mathrm{-index}}(t) = \mean{S\mathrm{-index}(t)} \mathrm{,}
\end{aligned}
\end{equation}
and $\epsilon(t)$ are white noise models containing measurement noise and jitter terms. BIS is proved to be interlinked with both $G(t)$ and its derivative $\dot{G}(t)$, while S-index is exclusively related to $G(t)$ \citep[e.g.,][]{2010ApJ...725..875I,Dumusque14,2017MNRAS.468L..16T}. Hence we exclude $\dot{G}(t)$ term for $S$-index. In terms of choices of kernels for $G(t)$, we opted for MEP kernels available in \texttt{S+LEAF} introduced earlier, which is a faster approximation of the QP kernel. To explore the posterior of free parameters, we used nested sampling through \texttt{PolyChord}\footnote{\url{https://github.com/PolyChord/PolyChordLite}} \citep{2015MNRAS.450L..61H,2015MNRAS.453.4384H}, with all priors on the free parameters set as uninformative. This nested sampling approach is favoured here as it demonstrates better performance when compared to MCMC approaches, especially in datasets exhibiting multi-modality.

\begin{figure*}
	\centering
	\includegraphics[width=1.0\textwidth]{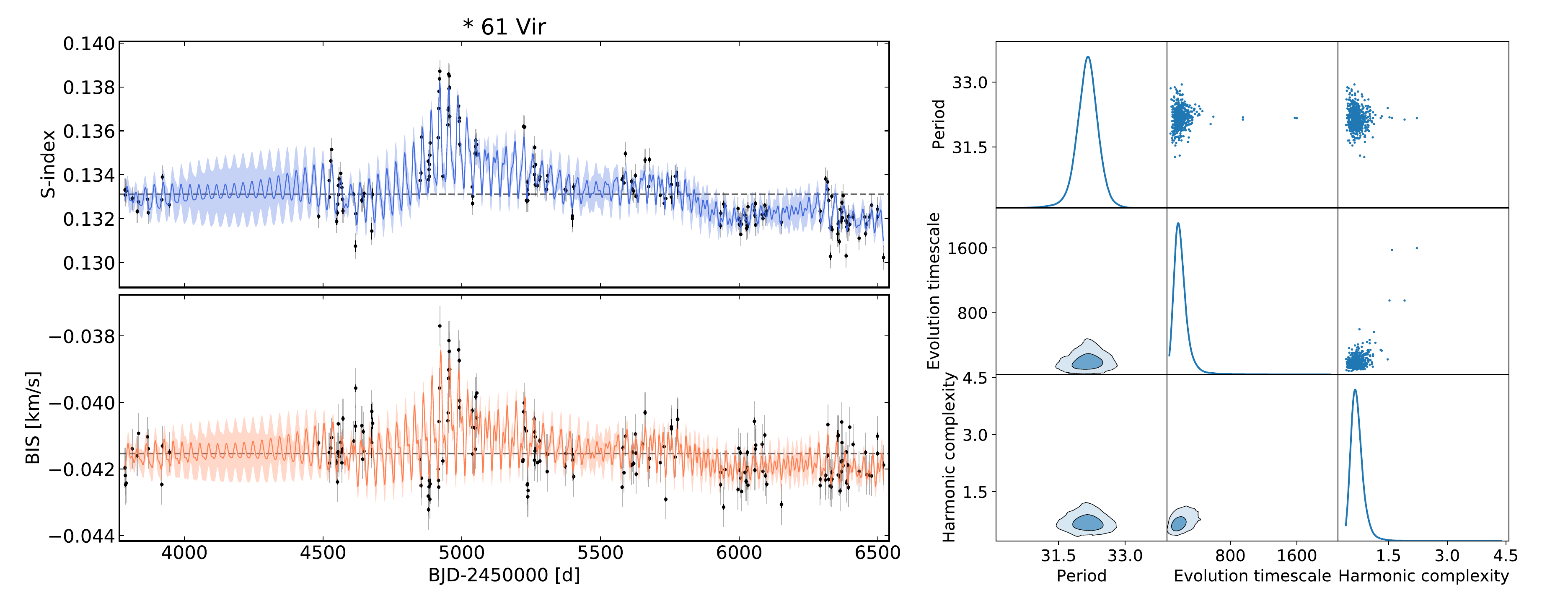}
    \caption{Demonstration of multi-dimensional GP fit on the * 61 Vir (HD~115617) dataset. In the left panel, the black markers show the time-series of $S$-index and BIS, and their associated uncertainties. The coloured (blue and coral) lines show the GP predictions over the observations, and the corresponding shaded areas of the lines show one $\sigma$ uncertainty of predicted distributions. The horizontal dashed grey lines show the means of the time-series. The grey extensions on the errorbars show the jitter terms. The right panel show posterior distributions of selected model parameters, including period, evolution timescale and harmonic complexity.}
    \label{fig:* 61 Vir}
\end{figure*}

We show an example of implementing the framework using the * 61 Vir dataset. In Table \ref{tab:* 61 Vir-param}, we show the inferred values of the free parameters in the model based on the posterior distributions, alongside the priors designated for the parameters during the sampling. The inferred values are the median of the posterior distributions, and the statistical uncertainties are given by the 16th to 84th percentile range.

\renewcommand{\arraystretch}{1.4}
\begin{table*}
	\centering
	\caption{Priors and inferred value of selected parameters in the model for the example on the * 61 Vir (HD~115617) dataset.}
	\label{tab:* 61 Vir-param}
	\begin{tabular}{lcr}
        \hline Parameter & Prior $^{*}$ & Inferred value \\
        \hline
        $\rm{Period}\ P \ (\mathrm{d})$ & $\mathcal{U}[0.1,60.0]$ & $32.1_{-0.2}^{+0.2}$ \\
        $\rm{Evolution \ timescale}\ l \ (\mathrm{d})$ & $\mathcal{U}[1,5000]$ & $186_{-50}^{+78}$ \\
        $\rm{Harmonic \ complexity}\ \Gamma $ & $\mathcal{U}[0.01,5.00]$ & $0.68_{-0.12}^{+0.16}$ \\
        $a_{\mathrm{BIS}}(\mathrm{~m} \mathrm{~s}^{-1})$ & $\mathcal{U}[\rm{-rms(BIS)},\rm{rms(BIS)}]$ & $0.55_{-0.08}^{+0.15}$ \\
        $b_{\mathrm{BIS}}(\mathrm{~m} \mathrm{~s}^{-1} \mathrm{~d}^{-1})$ & $\mathcal{U}[\rm{-10*rms(BIS)},\rm{10*rms(BIS)}]$ & $-3.5_{-0.9}^{+0.7}$ \\
        $a_{S\mathrm{-index}}(\mathrm{~d}^{-1})$ & $\mathcal{U}[\rm{-rms(S-index)},\rm{rms(S-index)}]$ & $0.0013_{-0.0002}^{+0.0003}$ \\
        \hline
    \end{tabular}
    {\\ \footnotesize * $\mathcal{U}$[a,b] refers to a uniform prior between a and b.}
\end{table*}

\renewcommand{\arraystretch}{1.0}

Figure \ref{fig:* 61 Vir} shows a multi-dimensional GP fits to the time-series of * 61 Vir (HD~115617) dataset. In the left panel, the black markers show the time-series of $S$-index and BIS, and their associated uncertainties. The coloured (blue and coral) lines show the GP predictions over the observations, and the corresponding shaded areas of the lines show one $\sigma$ uncertainty of predicted distributions. The horizontal dashed grey lines show the means of the time-series. The grey extensions on the errorbars show the jitter terms. The right panel show posterior distributions of selected model parameters, including period, evolution timescale and harmonic complexity.

\section{Stellar rotation} \label{sec:rot}
\subsection{Rotation period measurement} \label{sec:prot}

\begin{figure}
	\centering
	\includegraphics[width=1.0\columnwidth]{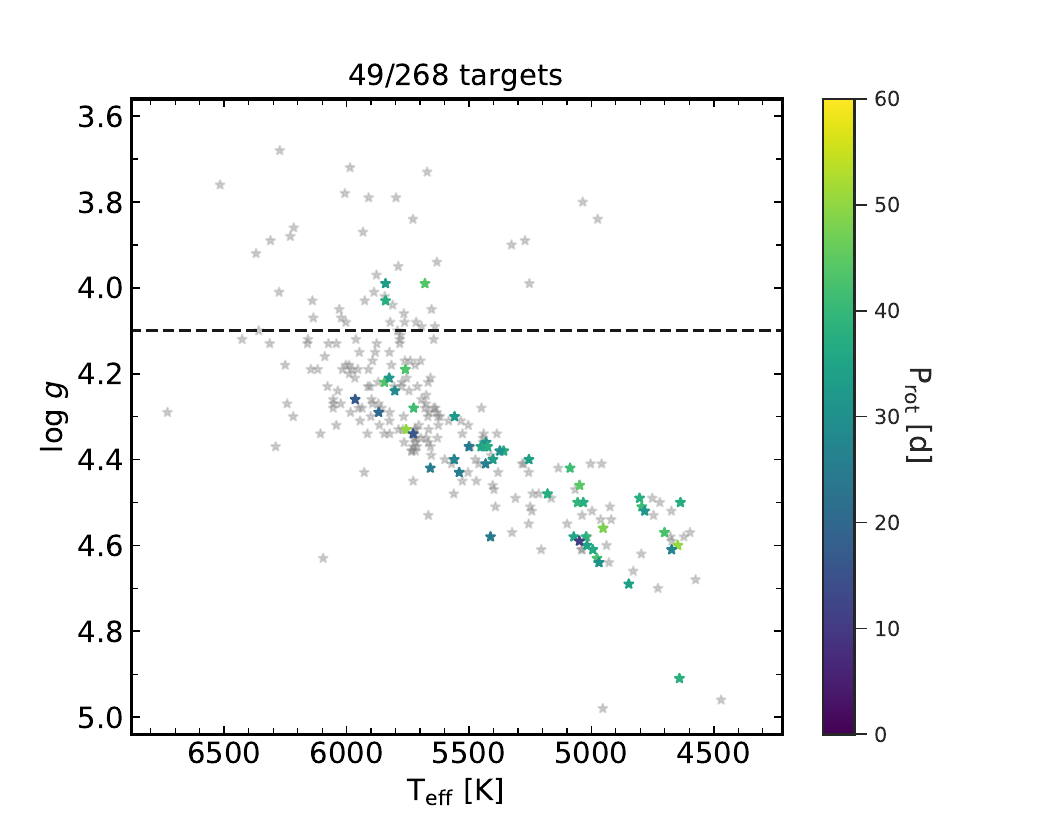}
    \caption{HR diagram of our sample with the coloured points showing the measured rotation period for the 49 targets which passed our vetting procedure. The grey markers denote the other targets for which the rotation periods were not recovered or were not considered sufficiently robust to include in subsequent analysis. The dotted line at $\rm{log} \ g = 4.1$ approximately separates main-sequence stars from evolved stars, i.e., sub-giants.}
    \label{fig:prot}
\end{figure}

The variations induced by active regions in activity indicators, modulated by the rotation of the star, are believed to closely represent the star's true rotation period \citep[e.g.,][]{Angus18,Nicholson2022}. As such, the rotation period is captured by the hyper-parameter 'period' in our multi-dimensional GP model when the activity-induced signal is well-modelled.

We applied the multi-dimensional GP framework with a MEP kernel to each star in our sample. To determine whether the period is securely detected, on a separate analysis, we applied the same framework but used a Matern 3/2 kernel instead with GP, representing an aperiodic model in comparison to the above periodic model (with the MEP kernel). The evidence $Z$ of each model is estimated through nested sampling to calculate $\rm{\Delta ln} \ Z = \rm{ln} \ Z_{periodic} - \rm{ln} \ Z_{aperiodic}$. A positive value indicated that the periodic model is preferred. We then used $\Delta \rm{ln} \ Z$ as our starting criterion to select targets where the periodic model is preferred. Additionally, we rejected cases through visual inspection where the $\Delta \rm{ln} \ Z$ was unreliable because of two possible scenarios. Firstly, in cases where neither model fits well, the $\Delta \rm{ln} \ Z$ criterion could be biased as the Bayesian model comparison is only meaningful when both models adequately fit the data. Secondly, the MEP kernel (or the quasi-periodic kernel in general) we employed has regions in the parameter space that could behave aperiodically. In such cases, even if the periodic model appears preferable based on evidence difference, secure detection of the period is not guaranteed.
Upon thorough vetting, we identified 49 targets for which we believe the model adequately captured the activity-induced variations, resulting in an accurate recovery of their true rotation periods. 

We also want to emphasize that in comparison to conventional methods, such as periodograms or the autocorrelation function (ACF), the GP approach displays greater resilience against harmonic misdetection when determining rotation periods \citep[e.g.,][]{Angus18}. This robustness arises from the utilization of the MEP or QP kernel, where potential harmonic components are inherently considered, i.e., the 'harmonic complexity' hyperparameter serves as an approximate gauge of the harmonic contribution. This ensures the algorithm finds the true rotation period which is associated with the lowest base frequency.

The selected 49 targets are shown in an HR diagram in Figure \ref{fig:prot}, with the colour of the markers representing their recovered rotation periods. All other targets of which the periods have not been successfully recovered are denoted with grey markers. The dotted line at $\rm{log} \ g = 4.1$ approximately separates main-sequence stars from evolved stars, i.e., sub-giants. Notably, we find that the rotation period broadly decreases as effective temperature rises (or for earlier spectral types), although there is a considerable amount of scatter at any given temperature. This is broadly consistent with both theoretical expectations \citep[e.g.,][]{1972ApJ...171..565S} and preceding rotation period surveys \citep[e.g.,][]{MQ14}. Further insights on age-activity-rotation relations are discussed in section \ref{sec:spin-down}. 

Regarding the detection rate, there are approximately 23\% of main-sequence G and K stars of which the rotation period is well recovered. However, this rate decreases significantly for earlier type stars such as F stars, and for sub-giants. The rapid rotation of F stars, typically with periods under 10 days, necessitates denser time sampling than what is currently available to accurately determine their rotation periods. In the case of sub-giants, their evolving nature results in diminished or altered activity patterns, deviating from the assumption of our activity model and thus reducing the detection rate. It is important to note that the detection rate for the rotation period can be considered as an indirect measure of the model's efficacy at mitigating activity signals across different stellar types. In this context, the model demonstrates the best performance for G and K stars. For F stars, the model's efficacy could be enhanced with optimized sampling, tailored to capture their rapid rotation. However, it is essential to recognize that effective activity modelling does not necessarily equate to a target's appropriateness for planet detection. For example, challenges in modelling activity could arise due to a lack of pronounced active regions on the stellar surface. In such cases, the amplitude of the activity signal remains minimal and as such the star could still possibly be an attractive candidate for exoplanet detection.

Last but not least, it is important to highlight that compared to traditional methods like the Lomb-Scargle periodogram, the GP method holds an advantage in determining the signal's period when the rotational modulation signal evolves over time \citep{Angus18}. Additionally, with the multi-GP method used in this work, we search for and identify common periodic signatures among multiple time-series, thereby increasing the confidence in the measured rotation period. Moreover, the multi-GP framework offers insights beyond just the rotation periods of the stars. Specifically, it provides a quantitative understanding of the covariance structure of the activity signals. Such information can be integrated into statistical frameworks to disentangle activity from planetary signals in the RVs. This can be achieved through either simple linear regression models, or a more complex framework that explicitly takes the covariance structure into account, such as the L1 periodogram of \citet{Hara17}. We defer further exploration of this potential to future work.

\subsection{Rossby number} \label{sec:ro}

\begin{figure}
	\centering
	\includegraphics[width=1.0\columnwidth]{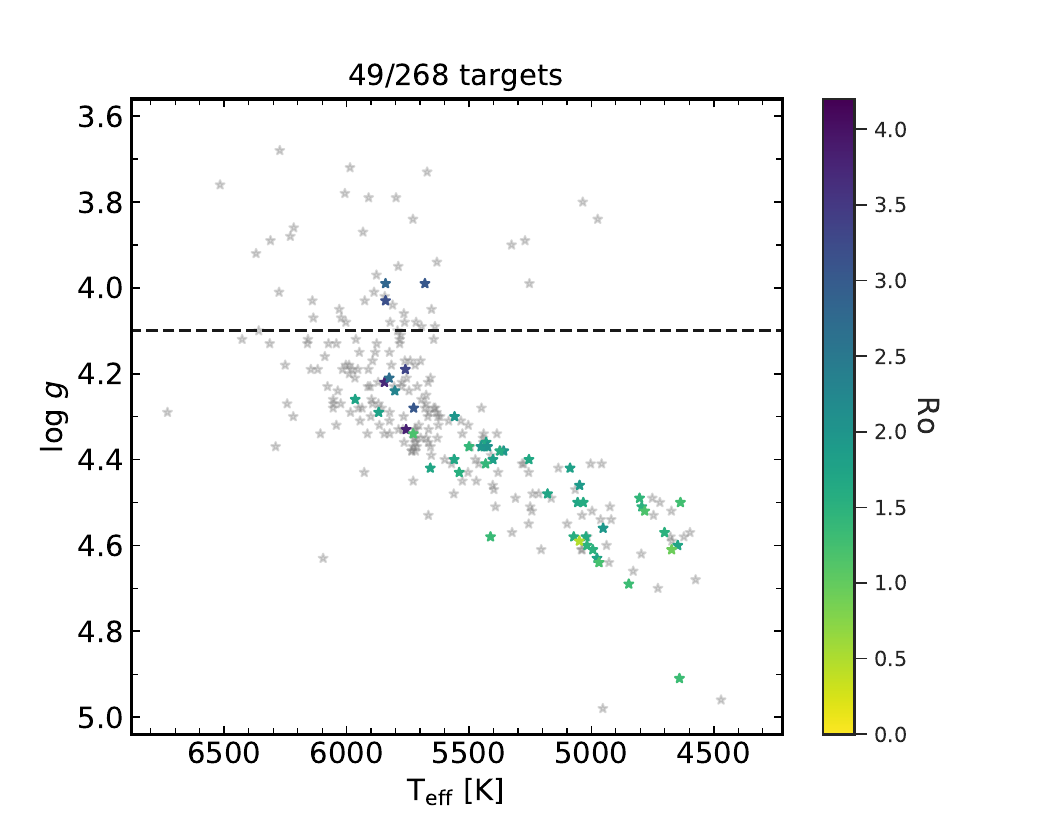}
    \caption{As Figure~\ref{fig:prot}, but the colour scale now shows the Rossby number.}
    \label{fig:ro}
\end{figure}

The Rossby number $\rm{Ro}$, which is given by the ratio of the rotation period $P_{\mathrm{rot}}$ to the convective turnover timescale $\tau_{\mathrm{c}}$:
\begin{equation}
\rm{Ro} = P_{\mathrm{rot}} / \tau_{\mathrm{c}},
\end{equation}
is a key parameter of the dynamo process that gives rise to large-scale magnetic fields among the stars we are considering in this study. Numerous empirical studies have shown that $\rm{Ro}$ is a key parameter for the angular momentum evolution, as well as for magnetic field strength and topography of late-type stars \citep[see e.g.][]{1984ApJ...279..763N}. To first order, it is an indicator of the activity level of the star: the lower the Rossby number, the more active the star. 

We evaluated the Rossby number for the stars in our sample with rotation period measurements, using the empirical relations from \citet{2011ApJ...741...54C} to calculate $\tau_{\mathrm{c}}$:
\begin{equation} \label{eqa:tc}
\begin{aligned}
\tau_{\mathrm{c}}\left(T_{\text {eff }}\right)= & 314.24 \exp \left[-\left(\frac{T_{\text {eff }}}{1952.5 \mathrm{~K}}\right)-\left(\frac{T_{\text {eff }}}{6250 \mathrm{~K}}\right)^{18}\right] \\
& +0.002 .
\end{aligned}
\end{equation}.

The Rossby number of the sample is shown in the right panel of Figure \ref{fig:ro}. The coloured points show the selected 49 targets from the visual vetting. Most of the stars in our sample have a relatively high Rossby number, i.e., ${\rm Ro}>1.0$, and are thus inactive stars. This is unsurprising, as most archival HARPS observations were taken as part of planet surveys, which tend to avoid active stars. Unlike $\log R_{\rm{HK}}^{\prime}$, the correlation between the Rossby number and the effective temperature is relatively weak, and can be explained mostly by the fact that the hotter stars in our sample are starting to evolve off the main sequence.

\subsection{Implications for spin-down of middle-aged stars} \label{sec:spin-down}
Sun-like stars are expected to undergo spin-down following a simple power law \citep[e.g.,][]{Skumanich1972} to the first order during their main sequence stage. This spin-down emerges from the continuous loss of angular momentum, a consequence of torques generated from interactions between the star's surface magnetic field and its stellar wind. However, subsequent observations from open clusters spanning a wide age range challenged the Skumanich-like spin-down. Specifically, these results have brought to light two phases in a star's evolution where the spin-down process appears to `stall'  \citep{2009ApJ...695..679M,2018ApJ...862...33A,2019ApJ...879..100D}. 

To account for these stalling effects, several scenarios have been proposed. The two main hypotheses are the core-envelope coupling \citep[e.g.,][]{2011MNRAS.416..447S,2015A&A...584A..30L,2020A&A...636A..76S,2021A&A...649A..96J} happening at a few Myrs to a Gyr, and the weakened magnetic breaking \citep[e.g.,][]{vS16} happening at a few Gyrs. The former suggests angular momentum can be transferred from the radiative core to the convective envelope due to internal differential rotation, attenuating the spin-down of the envelope. The latter indicates that magnetic braking is substantially weakened once the star achieves a critical Rossby number, possibly due to a change in the magnetic field's morphology. This alteration -- shifting from a dipole field to a quadrupole field or an even more intricate configuration -- makes it less effective at shedding angular momentum \citep{2015ApJ...814...99R}.

Gaps in the rotation period versus effective temperature diagram have been found with Kepler stars \citep[e.g.,][]{2013MNRAS.432.1203M, MQ14,2021ApJ...913...70G}. These gaps can potentially be explained by the aforementioned stalling effects. \citet{David22} provide stronger evidence by combining the \emph{Kepler} rotation periods with more accurate spectroscopic effective temperatures from, e.g., the Large Sky Area Multi-Object Fiber Spectroscopic Telescope (LAMOST) \citep{2012RAA....12.1197C,2012RAA....12..723Z}. Their findings indicate two prominent pile-ups in the period--effective temperature diagram. The short-period pile-up is attributed to the core-envelope coupling mechanism, while the long-period pile-up is linked to the phenomenon of weakened magnetic braking. This latter association is primarily because it aligns seamlessly with a consistent Rossby number curve.

\begin{figure*}
	\centering
	\includegraphics[width=0.72\textwidth]{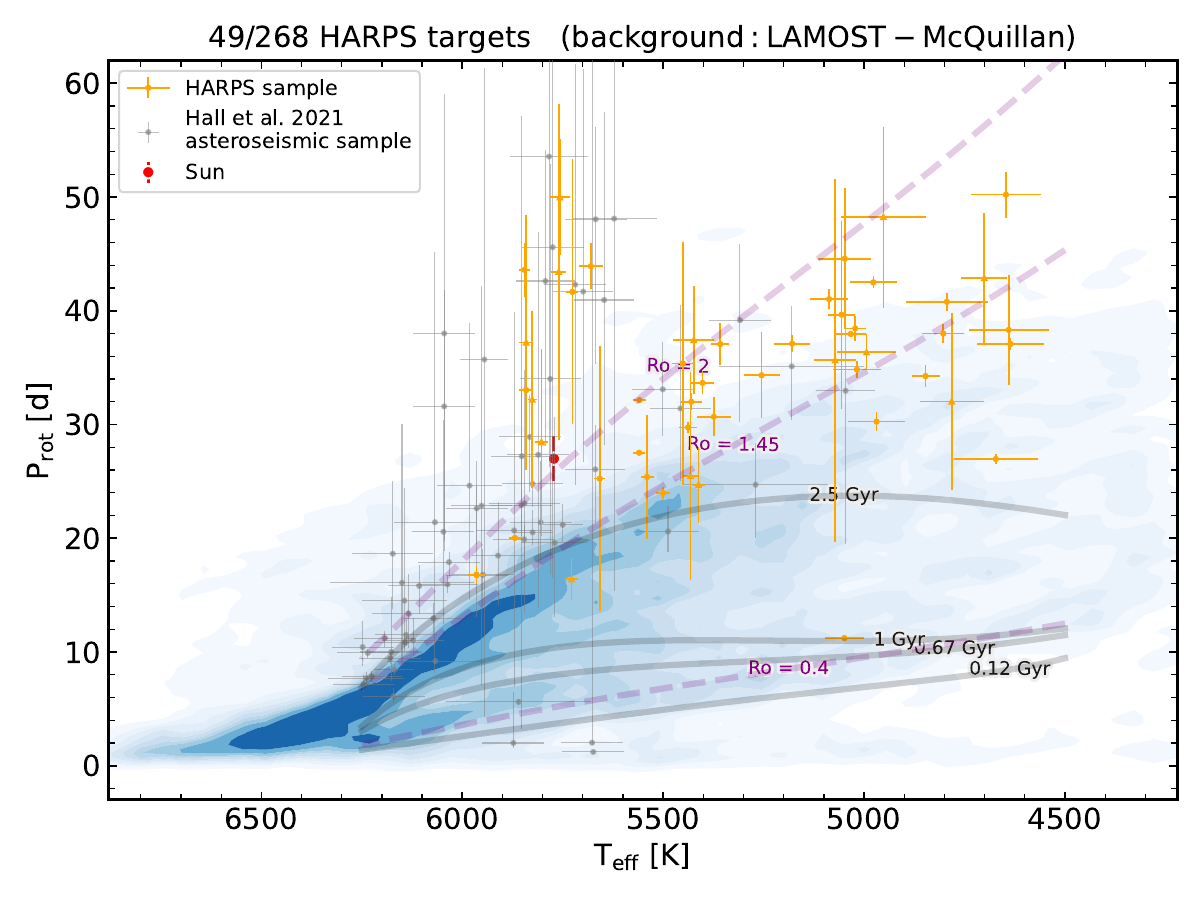}
    \caption{Rotation period versus effective temperature for our sample, compared to previous results from the literature. The blue shading in the background represents a kernel density estimate of the \emph{Kepler}/LAMOST period-temperature results. Empirical cluster sequences of different ages are shown in grey lines, derived from \citet{2020ApJ...904..140C}. Lines of constant Rossby number are shown in purple. The asteroseismic sample of \citet{2021NatAs...5..707H} is plotted in grey and our new results based on HARPS data are plotted in orange.}
    \label{fig:prot-comp1}
\end{figure*}

We constructed a similar diagram, by combining rotation periods measured in \emph{Kepler} data by \citet{MQ14} with effective temperatures from LAMOST. This resulted in a blue Kernel Density Estimation (KDE) plot, displayed in the background of Figure \ref{fig:prot-comp1}, generated with the codes \footnote{\url{https://github.com/trevordavid/rossby-ridge}} provided by \citet{David22}. Empirical cluster sequences of different ages are shown in grey lines, derived from \citet{2020ApJ...904..140C}. The purple lines correspond to constant Rossby numbers, following
\begin{equation}
P_{\rm{rot}} \left(\rm{Ro}, T_{\rm{eff }}\right) = \rm{Ro} \times \tau_{\rm{c}} \left(T_{\rm{eff}} \right),
\end{equation}
with convective turnover timescales from \citet{2011ApJ...741...54C} as Equation \ref{eqa:tc}.

In the KDE, the long-period pile-up is visible below the ${\rm Ro}=1.45$ curve, while the short-period pile-up appears above the ${\rm Ro}=0.4$ curve. The Sun is shown as a reference in red, with $T_{\text{eff},\odot} = 5772 \ \rm{K} $ \citep{2016AJ....152...41P} and $P_{\text{rot},\odot} = 27 \pm 2 \ \rm{d}$ estimated from \citet{1990ApJ...351..309S}. We have overlaid the asteroseismic sample from \citet{2021NatAs...5..707H}, depicted in grey, and our HARPS sample of 49 well-determined rotation periods (detailed in section \ref{sec:prot}) in orange. Both samples clearly occupy a distinct period-temperature space to the much larger \emph{Kepler}, predominantly located above the long-period pile-up. We interpret this primarily as a consequence of the different selection effects for the three samples. The \emph{Kepler} sample, while much larger and benefiting from tight, regular time sampling, relies on detecting the rotational modulation of active regions in broad-band optical light curves. This becomes increasingly difficult for less active stars as the fraction of the stellar disk covered by active regions drops, and for slower rotators as the rotation period becomes comparable to both the lifetime of the active regions and the duration of individual \emph{Kepler} `quarters'. The asteroseismic sample uses $p$-modes to measure internal and surface rotation rates, and is strongly biased towards hotter and lower gravity (evolved) stars, for which the amplitude of the $p$-modes is larger and their frequencies lower than for later type main-sequence stars. The HARPS sample spans a longer baseline and uses spectroscopic activity indicators rather than broad-band photometry to detect the rotational modulation of active regions. It is reasonable to expect that targetted indicators should be more sensitive to active regions than photometry, particularly as the active regions decrease in size and number and the rotation slows. A possible additional contributing factor is the fact that, while dark spots dominate the activity patterns for active stars, less active stars tend to be more faculae-dominated \citep{Meunier2010}. The photometric signatures of faculae are rather subtle, whereas they have strong chromospheric signatures, which are probed by spectroscopic indicators such as $\log R'_{\rm HK}$ and also have a strong signature in line-shape indicators such as the BIS via their local dampening effect on convective flows. A more in-depth discussion on this topic can be found in section \ref{sec:f-to-s}.

\begin{figure*}
	\centering
	\hspace*{+2.4cm}\includegraphics[width=0.8\textwidth]{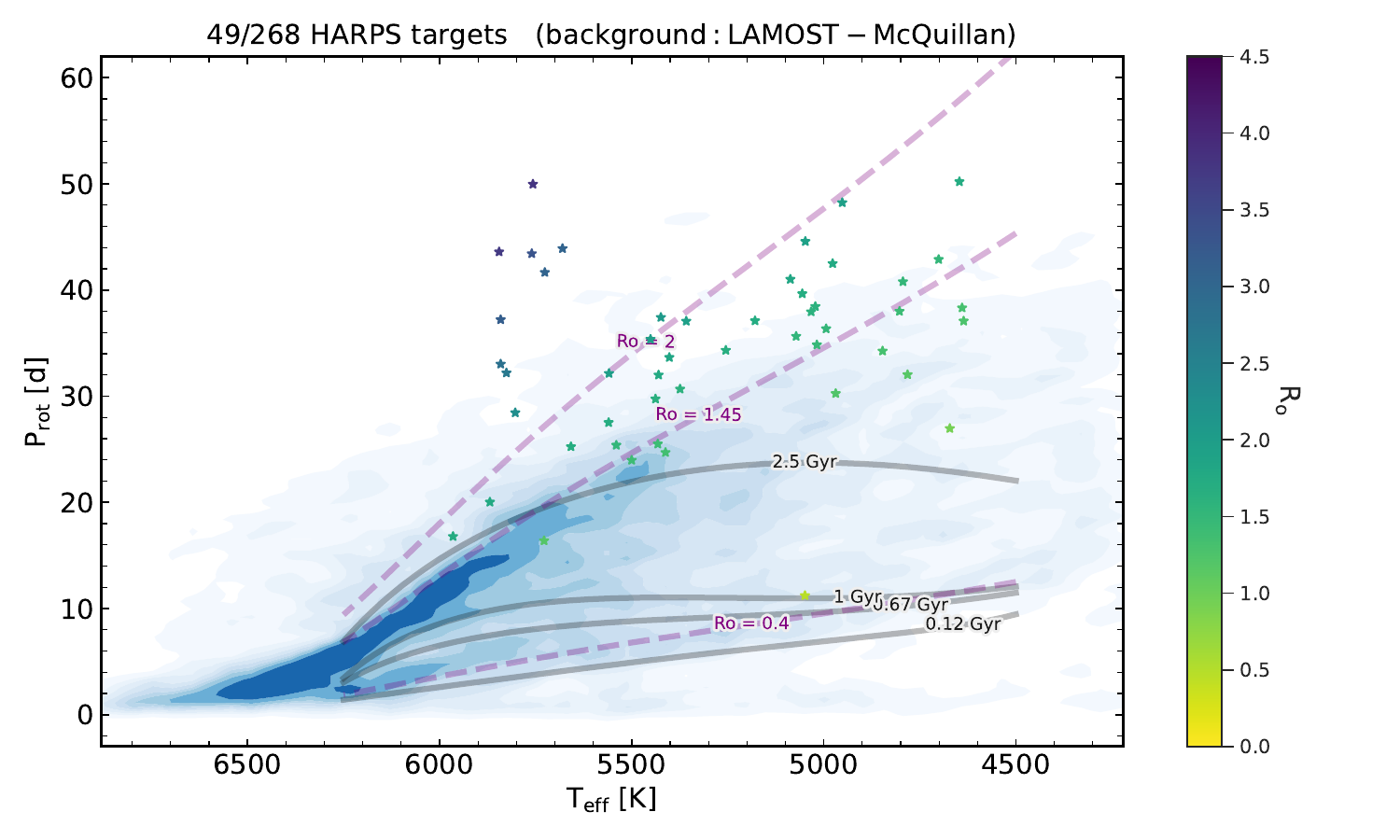}
    \caption{As Figure~\ref{fig:prot-comp1}, but the points now show our sample colour-coded according to Rossby number.}
    \label{fig:prot-comp2}
\end{figure*}

Considering the combined HARPS and asteroseismic samples, we see that two distinct populations emerge. The first is composed of hotter stars (with $T_{\rm eff} \ge 5700$\,K), mostly from the asteroseismic sample, though with a handful of HARPS measurements. This population follows a very steep period-temperature relation, consistent with the rapid spin-down expected as stars leave the main sequence and start to evolve towards the red giant branch.

The second population, predominantly from our sample, consists of cooler (mid-G to mid-K) main-sequence stars. It follows a shallower period-temperature relation, and its upper envelope corresponds to the ${\rm Ro}=2$ curve.
This second population could possibly be the extension of the long-period pile-up observed in the \emph{Kepler} sample, but skewed towards cooler stars. In the \emph{Kepler} sample, the long-period pile-up is less pronounced at cooler temperatures, but this could be a selection effect: at lower temperatures, the pile-up moves to longer periods, which correspond to lower photometric amplitudes, and the stars also become intrinsically fainter.
In Figure \ref{fig:prot-comp2}, we show the Rossby numbers computed for our sample compared to the KDE of the \emph{Kepler} sample. They fall in the range of 1.5--2.5, consistent with the critical Rossby number inferred from the photometric long-period pile-up. A slight discrepancy between this tail and the anticipated long-period pile-up extension might arise from biases in effective temperature measurements from different sources, as argued in \citet{David22} in the context of the asteroseismic sample.

\section{Faculae-to-spots ratio} \label{sec:f-to-s}

\begin{figure*}
	\centering
	\includegraphics[width=1.0\textwidth]{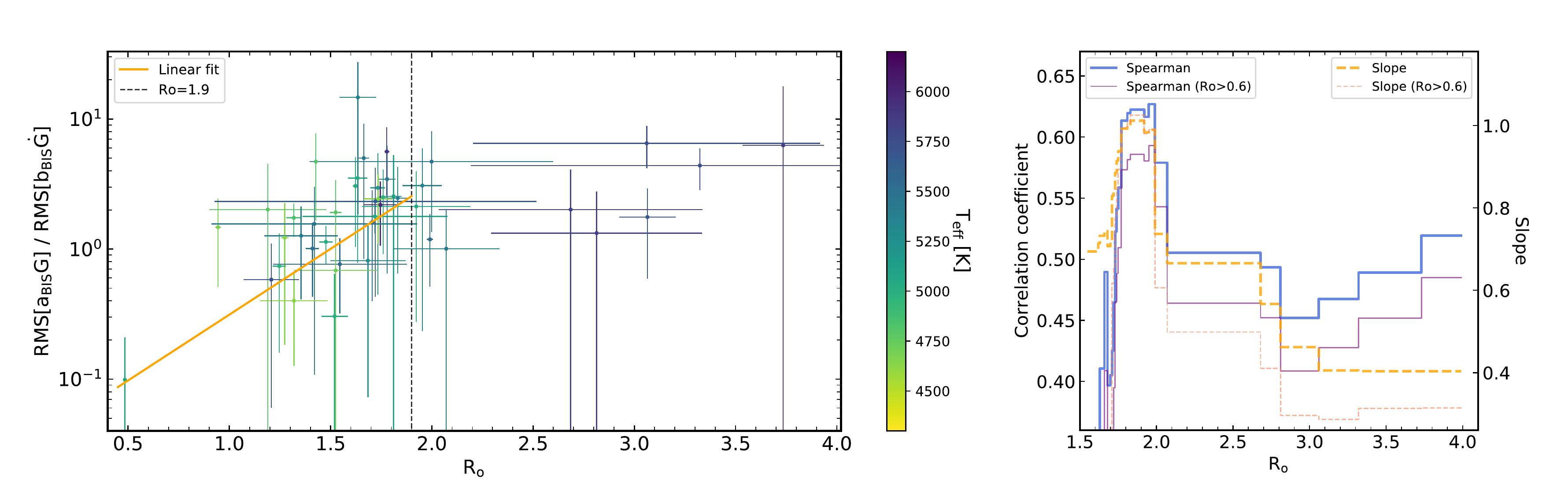}
    \caption{The left panel shows the RMS ratio $A/B$ (a proxy for the faculae-to-spot ratio) versus the Rossby number ${\rm Ro}$ (more magnetically active stars have smaller Ro). The error bars show the derived statistical uncertainties, and the colour scheme indicates the effective temperature. The dotted grey line indicates ${\rm Ro} = 1.9$, and the orange line shows the linear fit to the data within the range of ${\rm Ro}<1.9$. The right panel shows the estimated Spearman correlation coefficient between $A/B$ and ${\rm Ro}$, as well as the slope of the linear fit to $\rm{log} (A/B)$ as a function of ${\rm Ro}$. The two proxies are plotted against the cut-off value of ${\rm Ro}$ ranging from 1.6 to 4.1. The blue and orange lines show the analysis of the full data, while the thinner purple and coral lines show the analysis excluding the data point at ${\rm Ro} \approx 0.5$ for comparison.}
    \label{fig:f/s}
\end{figure*}

Spots and faculae (or their chromospheric counterpart, plages) are the two main types of active regions that are expected to induce variations in the RVs and activity indicators used in planet searches. These variations are induced by two simultaneous processes. The first, known as the photometric effect, is thought to be mainly induced by dark spots, which locally reduce (or even suppress entirely) the local flux emerging from the stellar surface. The other is due to the fact that convective up-flows are suppressed in regions of enhanced magnetic flux density, which includes both spots and faculae. This results in a local reduction in the net blue-shift caused by convective flows relative to the "clear" photosphere, known as the Inhibition of Convective Blueshift (ICB) effect. Both processes cause distortions to the disk-integrated line profiles, resulting in variations the RV, FWHM and BIS, while chromospheric activity indicators such as the $S$-index or $\log R'_{\rm HK}$ are sensitive to the fractional coverage of the active regions themselves.

The differentiating factor between spots and faculae lies in the processes through which they induce variations. While spots cause variations via both these processes, faculae primarily influence the ICB effect \citep[e.g.,][]{Dumusque14}. Within the context of our multi-dimensional GP framework, the terms $a G(t)$ and $b \dot{G}(t)$ distinctively represent these two effects \citep[e.g.,][]{A12,V15}. Hence, the ratio of the RMS values of these two terms, $\mathrm{{RMS}[}a G(t)\mathrm{]} \ / \ \mathrm{{RMS}[}b \dot{G}(t) \mathrm{]}$, which we denote as $A/B$ in the following text, can be interpreted as an approximate representation of the relative faculae-to-spots ratio. 
This potentially offers an interesting way to measure the faculae-to-spots ratio in the sample of stars where we consider the results of the GP modelling to be sufficiently robust.

In turn, doing this would allow us, in principle, to test the widely held paradigm that more magnetically active stars are more spot-dominated, while quieter stars have fewer spots and more faculae \citep[e.g.,][]{2010A&A...512A..39M,2020A&A...642A.225A}. Of course, direct measurements of the star's magnetic field necessitate polarimetric observations, and few stars in our sample have such observations. However, the Rossby number ${\rm Ro}$ derived in Section \ref{sec:ro} is known to correlate strongly with the strength of the magnetic field $\rm{<B>}$ \citep[e.g.,][]{Reiners2014,Vidotto2014,2022A&A...662A..41R}, with an exception for very inactive stars (i.e., ${\rm Ro} > 2$).

For the targets of which we have detected their rotation periods securely, we examined the relationship between the RMS ratio $A/B$ and the derived Rossby number ${\rm Ro}$ in the left panel of Figure \ref{fig:f/s}. The error bars show the propagated statistical uncertainties, and the colour scheme indicates the effective temperature of the stars. We excluded samples exhibiting uncertainties in the $A/B$ exceeding 200$\%$.

To evaluate the monotonic relationship between the $A/B$ and the ${\rm Ro}$, we calculated the Spearman correlation coefficient, considering an upper cut-off value for ${\rm Ro}$ in the range of 1.6 to 4.0. The result is shown in the right panel of Figure \ref{fig:f/s}. We found that the Spearman coefficient peaks between $1.8<{\rm Ro}<2.0$ with a value of 0.62, after which the value drops rapidly. In addition, we conducted linear fits to $\rm{log}(A/B)$ as a function of ${\rm Ro}$, and showed the slope of the fit versus the ${\rm Ro}$ cut-off value in the same panel. We found that the slope also peaks between $1.8<{\rm Ro}<2.0$, behaving similarly to the Spearman coefficient. Both proxies indicate a relatively strong positive correlation between $A/B$ and ${\rm Ro}$ until $\rm Ro \approx 1.9$, after which the correlation significantly weakens.

Additionally, we conducted analyses to assess the impact of the data point at ${\rm Ro} \approx 0.5$ by excluding it from the correlation analysis. The results are represented as thinner lines in the same panel, labelled as ${\rm Ro}>0.6$. Both the Spearman coefficient and the slope showed behaviors remarkably similar to those observed in the full analysis. This similarity suggests that the overall results are not predominantly influenced by this single data point. Despite this, we assert that the detection at ${\rm Ro} \approx 0.5$ is robust. The lack of data in the $0.5<{\rm Ro}<1.0$ range is likely attributed to the HARPS sample's bias towards inactive stars.

The relationship between $A/B$ and ${\rm Ro}$ is aligned with theoretical predictions: a lower ${\rm Ro}$ implies a stronger magnetic field and consequently, a star that is more spot-dominated. This leads to a decreased facular-to-spot ratio, mirrored by a lower $A/B$, which is exactly the trend we have seen in the ${\rm Ro}<1.9$ regime. However, when ${\rm Ro}>1.9$, the correlation disappears, and there is no longer any obvious relationship between the $A/B$ and the Rossby number. We speculate that the disappearance of the correlation might arise from a morphological transition in the stellar magnetic field beyond ${\rm Ro} \approx 1.9$, away from a predominantly dipolar configuration. If the $B$-field structure becomes more complex, the correlation between $B$-field strength and Rossby number would not be expected to persist. 

Intriguingly, the Rossby number at which the correlation with the $A/B$ disappears,  ${\rm Ro} \approx 1.9$, is close to the ${\rm Ro}$ that marks the upper envelope of the \emph{Kepler} main-sequence rotation period-temperature distribution in Figures~\ref{fig:prot-comp1} and \ref{fig:prot-comp2}. It also corresponds approximately to the critical Rossby number in the range $1.4 \le {\rm Ro} \le 2.0$ postulated in stellar spin-down theories \citep[e.g.,][]{vS16,vS19,David22}, where the efficiency of stellar angular momentum dissipation is expected to drop significantly due to the same alterations in magnetic field morphology. 

A more detailed investigation of the faculae-to-spot ratio and its dependence on magnetic field strength would certainly be valuable, but would likely require more sophisticated activity indicators that can help disentangle the contributions of the different types of active regions (see e.g.\ \citealt{Cret_in_prep}) as well as a larger sample of stars with direct, polarimetric magnetic field measurements, and is therefore beyond the scope of this work.

\section{Conclusions} \label{sec:sum}
This paper presents the first results from an ongoing study of the activity properties of F, G and K stars using archival HARPS data. We applied the multi-dimensional GP framework developed by \citet{V15} and \citet{B22} for activity mitigation in RV planet searches to activity indicators extracted from HARPS spectra of 268 well-observed targets with precisely determined stellar parameters.
Applying the GP framework to such a large dataset was made possible by using the efficient implementation of our framework provided by \citet{D22}. While we do plan to perform a joint analysis of the activity indicators and the RVs in the future, this is made challenging by the presence in the RV time-series of residual systematic effects and an unknown number of planetary signals. A new version of the HARPS Data Reduction Software is under development (Dumusque et al.\ priv.\ comm.), which is expected to lead to significantly reduced systematics and has already allowed new planet candidates to be identified (with further HARPS observations being gathered to confirm them, ESO program 1110.C-4043, PI Hara). In the mean time, applying the multi-dimensional GP framework to the activity indicators alone already reveals a wealth of useful information. Our key findings are as follows.

We successfully recovered rotation periods for 49 slow rotators in our sample. We found that the rotation period decreases as effective temperature rises, i.e.\ for earlier spectral types, which is broadly consistent with both theoretical expectations \citep[e.g.,][]{1972ApJ...171..565S} and preceding rotation period surveys \citep[e.g.,][]{MQ14}. One limitation of our approach is that we do not consider the effects of differential rotation, primarily due to the sparse sampling of the majority of the targets in our sample. A possible extension of this work would be to analyse the data season by season and investigate seasonal variations in the period as well as in the other parameters of the model.

We placed our results in the context of existing, photometric estimates of field star rotation periods from $\emph{Kepler}$, and discussed their implications for the age-activity-rotation relations of F, G \& K stars. Our samples typically have longer periods than those derived from photometry, with the exception of the asteroseismic sample. We ascribe this to two factors. First, spectroscopic surveys like HARPS have a much longer baseline compared to photometric surveys. This allows for multi-seasonal monitoring, increasing sensitivity to slow rotators. Second, spectroscopic indicators are more sensitive to active regions, particularly the faculae that tend to dominate in slowly-rotating, magnetically inactive stars, than broad-band optical photometry.

Taken together with the pre-existing asteroseismic sample, our new sample of slowly-rotating ($P > \approx 30$\,d) FGK stars consists of two, distinct sub-populations. One is a potential extension of the long-period pileup towards cooler stars, with Rossby numbers in the range $1.4 < {\rm Ro}< 2.0$. The other consists of `hot slow rotators' with $T_{\rm eff} > 5700$\,K which we interpret as sub-giants which are rapidly spinning down as they start to expand. 

Overall, our results broadly agree with the findings of \citet{David22}, but this is the first time that the population of late G and K-type slow rotators is observed so clearly, above the `critical' Rossby number of ${\rm Ro} \approx 1.5$ that marks the upper envelope of the period-temperature distributions for \emph{Kepler} stars with photometrically derived surface rotation periods. We infer that angular momentum loss via a magnetised wind continues beyond this critical value, which instead marks the point where the active region covering fraction becomes too low to allow for photometric rotation period estimates. 

We also explored indirectly how the ratio of faculae to spots in active regions varies across our sample, using the RMS ratio of $a G(t)$ to $b \dot{G}(t)$ as a proxy. We find that this ratio is positively correlated with the Rossby number up to ${\rm Ro} \approx 1.9$, in accordance with theoretical expectations, since a lower ${\rm Ro}$ implies a stronger magnetic field and consequently, a more spot-dominated star. The correlation disappears for larger values of ${\rm Ro}$, indicating a possible transition in the stars' magnetic field morphology away from large-scale, predominantly dipolar fields that give rise to clearly detectable rotational modulation in the activity indicators. 

Thus, both our rotation period measurements and the relative contribution of higher-order versus lower-order behaviour in the activity indicator time-series are consistent with the idea that a significant transition in magnetic field morphology occurs around a critical Rossby number in the range $1.4 \le {\rm Ro} \le 2.0$. Notably, this transition aligns well with the critical ${\rm Ro}$ values suggested in stellar spin-down theories.

This paper demonstrates that rich information can be extracted from spectroscopic activity indicators, which sheds light on the structure and evolution of stars. Our intention is to extend this methodology to the RV once there is existing progress on the systemic correction and strategy for a thorough search of planetary signals. We note that the covariance structure of the activity signals learnt from this work can be effectively used to separate stellar and planetary components in the RVs.

The longer-term behaviour of the activity indicators analysed in this work can also be used to search for activity cycles in our targets. This will be the topic of the next paper in our series (Cr{\' e}tignier et al.\ in prep.). In the longer term, we also plan to incorporate the RV time-series, as well as other activity indicators, into our modelling, in order to improve our understanding of the strengths and limitations of the multi-dimensional GP framework for activity mitigation in RV planet searches, as a function of the stellar properties.

\section*{Acknowledgements}
H.Y., S.A., B.K., O.B., N.K.O.S. acknowledge funding from the European Research Council under the European Union’s Horizon 2020 research and innovation programme (grant agreement No 865624, GPRV).
N.K.O.S thanks the LSSTC Data Science Fellowship Program, which is funded by LSSTC, NSF Cybertraining Grant \#1829740, the Brinson Foundation, and the Moore Foundation; her participation in the program has benefited this work. M.C. acknowledges the SNSF support under grant P500PT\_211024.
Based on observations made with ESO Telescopes at the La Silla Paranal Observatory under the HARPS Guaranteed Time Observations (GTO) programme.
This research has made use of the SIMBAD database operated at CDS, Strasbourg (France).

\section*{Data Availability}

This work was based entirely on publicly available data download from the ESO archive. Tables \label{tab:star_prot} and \label{tab:star_all} are available in electronic format. 

\bibliographystyle{mnras}
\bibliography{ref}


\appendix

\section{Confirmed rotation period measurements}
Table \ref{tab:star_prot} shows the measured rotation periods $\rm{P_{rot}}$ for the selected 49 targets. In addition to $\rm{P_{rot}}$, we have provided information about the time span of the data $\rm{T_{span}}$, as well as all other hyperparameters from our model, including the evolution timescale $\rm{l}$ and the harmonic complexity $\rm{\Gamma}$. We have also included the derived Rossby number $\rm{Ro}$ and the $\mathrm{{RMS}[}a G(t)\mathrm{]} \ / \ \mathrm{{RMS}[}b \dot{G}(t) \mathrm{]}$ ($\rm{A/B}$; representing the facular-to-spot ratio). All values, except for $\rm{T_{span}}$, come with associated uncertainties.

\section{Basic properties of the sample}
Table \ref{tab:star_all} shows the basic properties of all 268 targets in the sample, including the effective temperature $T_{\rm{eff}}$, surface gravity $\rm{log} \ g$, and mean $\log R_{\rm{HK}}^{\prime}$ from the catalogue provided by \citet{GdS21}, with associated uncertainties. Additionally, we have included details about the temporal coverage of observations, denoted as $\rm{T_{range}}$, as well as the temporal span of the data processed using DRS 3.5. Any temporal span not covered by the DRS 3.5 processed data is filled by the DRS 3.8 processed data.

\onecolumn
\renewcommand{\arraystretch}{1.15}
\begin{longtable}{lccccccccccr}

    \caption{Confirmed rotation period measurements of 49 targets.}\label{tab:star_prot} \\
    \hline 
    Simbad name & $\rm{T_{span}}$ [yr] & $\rm{P_{rot}}$ [d] & $\rm{l}$ [d] & $\rm{\Gamma}$ & $\rm Ro$ & $\rm{A/B}$ \\
    \hline
    
BD-08 2823 & $2.42$ & $27.0_{-0.4}^{+0.4}$ & $966_{-462}^{+559}$ & $1.86_{-0.64}^{+0.91}$ & $0.94_{-0.02}^{+0.02}$ & $1.5_{-1.0}^{+1.0}$ \\
 HD 104067 & $2.38$ & $30.3_{-0.8}^{+0.8}$ & $386_{-230}^{+379}$ & $1.21_{-0.51}^{+0.66}$ & $1.25_{-0.03}^{+0.03}$ & $0.7_{-0.6}^{+0.6}$ \\
 HD 109200 & $7.24$ & $39.6_{-8.3}^{+8.3}$ & $57_{-38}^{+32}$ & $0.76_{-0.12}^{+0.34}$ & $1.72_{-0.36}^{+0.36}$ & $1.8_{-0.5}^{+0.5}$ \\
 HD 115617 & $7.49$ & $32.1_{-0.2}^{+0.2}$ & $186_{-50}^{+78}$ & $0.68_{-0.12}^{+0.16}$ & $1.99_{-0.02}^{+0.02}$ & $1.2_{-0.7}^{+0.7}$ \\
 HD 125072 & $7.44$ & $40.8_{-0.8}^{+0.8}$ & $482_{-227}^{+455}$ & $1.14_{-0.32}^{+0.47}$ & $1.52_{-0.03}^{+0.03}$ & $1.9_{-1.5}^{+1.5}$ \\
 HD 125184 & $9.38$ & $43.9_{-2.0}^{+2.0}$ & $200_{-78}^{+151}$ & $0.59_{-0.14}^{+0.19}$ & $3.06_{-0.14}^{+0.14}$ & $1.8_{-1.2}^{+1.2}$ \\
 HD 125595 & $3.44$ & $37.1_{-0.5}^{+0.5}$ & $804_{-410}^{+485}$ & $2.82_{-1.33}^{+1.78}$ & $1.27_{-0.02}^{+0.02}$ & $1.2_{-1.0}^{+1.0}$ \\
HD 13060 & $0.39$ & $34.3_{-3.8}^{+3.8}$ & $538_{-272}^{+420}$ & $3.91_{-1.75}^{+3.06}$ & $1.68_{-0.19}^{+0.19}$ & $0.8_{-0.7}^{+0.7}$ \\
 HD 136713 & $9.49$ & $36.4_{-1.6}^{+1.6}$ & $395_{-231}^{+527}$ & $0.87_{-0.35}^{+0.48}$ & $1.52_{-0.07}^{+0.07}$ & $0.3_{-0.3}^{+0.3}$ \\
HD 13808 & $8.08$ & $38.0_{-0.3}^{+0.3}$ & $379_{-106}^{+197}$ & $1.20_{-0.20}^{+0.28}$ & $1.62_{-0.01}^{+0.01}$ & $3.1_{-2.0}^{+2.0}$ \\
 HD 144628 & $9.44$ & $38.5_{-1.1}^{+1.1}$ & $184_{-64}^{+135}$ & $1.59_{-0.38}^{+0.64}$ & $1.63_{-0.05}^{+0.05}$ & $3.5_{-2.7}^{+2.7}$ \\
 HD 15337 & $10.94$ & $37.1_{-0.8}^{+0.8}$ & $1330_{-513}^{+456}$ & $5.43_{-2.48}^{+2.72}$ & $1.73_{-0.04}^{+0.04}$ & $3.0_{-2.5}^{+2.5}$ \\
 HD 154088 & $5.25$ & $30.7_{-1.7}^{+1.7}$ & $150_{-38}^{+66}$ & $1.53_{-0.35}^{+0.54}$ & $1.64_{-0.09}^{+0.09}$ & $14.7_{-12.8}^{+12.8}$ \\
 HD 154577 & $7.44$ & $34.3_{-1.0}^{+1.0}$ & $104_{-30}^{+42}$ & $1.00_{-0.14}^{+0.17}$ & $1.32_{-0.04}^{+0.04}$ & $1.7_{-0.5}^{+0.5}$ \\
HD 157172 & $5.40$ & $35.4_{-10.6}^{+10.6}$ & $498_{-247}^{+478}$ & $4.23_{-2.16}^{+2.96}$ & $2.00_{-0.60}^{+0.60}$ & $4.7_{-3.4}^{+3.4}$ \\
HD 157830 & $2.56$ & $25.4_{-5.4}^{+5.4}$ & $566_{-332}^{+722}$ & $3.46_{-1.48}^{+2.45}$ & $1.55_{-0.33}^{+0.33}$ & $0.8_{-0.4}^{+0.4}$ \\
 HD 161098 & $9.43$ & $27.5_{-0.2}^{+0.2}$ & $323_{-94}^{+163}$ & $1.00_{-0.22}^{+0.37}$ & $1.71_{-0.01}^{+0.01}$ & $1.6_{-1.2}^{+1.2}$ \\
 HD 16417 & $3.33$ & $33.0_{-6.1}^{+6.1}$ & $102_{-38}^{+90}$ & $0.83_{-0.23}^{+0.51}$ & $2.81_{-0.52}^{+0.52}$ & $1.3_{-1.4}^{+1.4}$ \\
 HD 168863 & $2.36$ & $32.0_{-7.8}^{+7.8}$ & $1000_{-529}^{+605}$ & $2.58_{-1.28}^{+2.71}$ & $1.19_{-0.29}^{+0.29}$ & $2.0_{-2.5}^{+2.5}$ \\
 HD 171587 & $2.56$ & $24.7_{-3.3}^{+3.3}$ & $650_{-390}^{+583}$ & $2.76_{-1.18}^{+1.95}$ & $1.35_{-0.18}^{+0.18}$ & $1.3_{-0.9}^{+0.9}$ \\
HD 172513 & $1.18$ & $24.0_{-0.6}^{+0.6}$ & $276_{-127}^{+173}$ & $1.14_{-0.34}^{+0.40}$ & $1.41_{-0.03}^{+0.03}$ & $1.0_{-0.6}^{+0.6}$ \\
 HD 176986 & $10.06$ & $34.8_{-0.8}^{+0.8}$ & $95_{-24}^{+34}$ & $0.65_{-0.11}^{+0.12}$ & $1.48_{-0.03}^{+0.03}$ & $1.1_{-0.4}^{+0.4}$ \\
 HD 183658 & $3.40$ & $28.4_{-8.2}^{+8.2}$ & $803_{-460}^{+535}$ & $3.57_{-2.03}^{+2.85}$ & $2.300_{-0.66}^{+0.66}$ &  -- \\
 HD 189567 & $9.04$ & $41.7_{-11.7}^{+11.7}$ & $75_{-62}^{+52}$ & $0.64_{-0.11}^{+0.89}$ & $3.06_{-0.86}^{+0.86}$ & $6.5_{-2.3}^{+2.3}$ \\
 HD 192310 & $7.23$ & $41.0_{-0.9}^{+0.9}$ & $222_{-64}^{+175}$ & $1.23_{-0.21}^{+0.37}$ & $1.81_{-0.04}^{+0.04}$ & $2.5_{-2.8}^{+2.8}$ \\
HD 202206 & $1.29$ & $50.0_{-5.1}^{+5.1}$ & $918_{-352}^{+406}$ & $5.54_{-2.04}^{+2.27}$ & $3.810_{-0.39}^{+0.39}$ &  -- \\
HD 208704 & $2.28$ & $32.2_{-7.8}^{+7.8}$ & $878_{-507}^{+615}$ & $3.10_{-1.47}^{+2.29}$ & $2.69_{-0.65}^{+0.65}$ & $2.0_{-2.1}^{+2.1}$ \\
 HD 215152 & $8.10$ & $38.0_{-0.8}^{+0.8}$ & $182_{-42}^{+70}$ & $2.59_{-0.59}^{+0.86}$ & $1.43_{-0.03}^{+0.03}$ & $4.7_{-3.1}^{+3.1}$ \\
 HD 216770 & $9.26$ & $37.4_{-4.7}^{+4.7}$ & $314_{-149}^{+221}$ & $1.57_{-0.51}^{+0.81}$ & $2.07_{-0.26}^{+0.26}$ & $1.0_{-1.0}^{+1.0}$ \\
 HD 21693 & $5.47$ & $32.0_{-0.7}^{+0.7}$ & $321_{-119}^{+294}$ & $0.84_{-0.17}^{+0.31}$ & $1.78_{-0.04}^{+0.04}$ & $3.4_{-2.8}^{+2.8}$ \\
 HD 21749 & $0.39$ & $38.3_{-4.9}^{+4.9}$ & $382_{-222}^{+353}$ & $1.45_{-0.47}^{+0.60}$ & $1.32_{-0.17}^{+0.17}$ & $0.4_{-0.3}^{+0.3}$ \\
 HD 22049 & $0.23$ & $11.2_{-0.1}^{+0.1}$ & $128_{-56}^{+74}$ & $1.70_{-0.58}^{+0.58}$ & $0.48_{-0.01}^{+0.01}$ & $0.1_{-0.1}^{+0.1}$ \\
 HD 223171 & $12.48$ & $37.2_{-11.2}^{+11.2}$ & $85_{-64}^{+163}$ & $0.98_{-0.38}^{+2.82}$ & $3.170_{-0.95}^{+0.95}$ &  -- \\
 HD 26965 & $12.41$ & $35.6_{-16.0}^{+16.0}$ & $51_{-40}^{+52}$ & $1.64_{-0.77}^{+3.74}$ & $1.560_{-0.70}^{+0.70}$ &  -- \\
 HD 27894 & $13.56$ & $48.2_{-8.0}^{+8.0}$ & $827_{-635}^{+762}$ & $3.34_{-1.60}^{+2.80}$ & $1.968_{-0.32}^{+0.32}$ &  -- \\
 HD 36003 & $11.21$ & $50.2_{-2.0}^{+2.0}$ & $214_{-78}^{+129}$ & $0.85_{-0.15}^{+0.22}$ & $1.74_{-0.07}^{+0.07}$ & $2.4_{-1.3}^{+1.3}$ \\
HD 40307 & $7.42$ & $42.5_{-0.5}^{+0.5}$ & $494_{-139}^{+273}$ & $1.21_{-0.20}^{+0.29}$ & $1.76_{-0.02}^{+0.02}$ & $2.5_{-1.6}^{+1.6}$ \\
 HD 45184 & $15.01$ & $20.0_{-0.1}^{+0.1}$ & $278_{-78}^{+115}$ & $1.22_{-0.21}^{+0.27}$ & $1.78_{-0.01}^{+0.01}$ & $5.6_{-3.0}^{+3.0}$ \\
 HD 4915 & $13.33$ & $25.2_{-11.7}^{+11.7}$ & $180_{-163}^{+326}$ & $2.49_{-1.28}^{+2.46}$ & $1.72_{-0.80}^{+0.80}$ & $2.3_{-1.9}^{+1.9}$ \\
 HD 51608 & $7.43$ & $37.1_{-1.8}^{+1.8}$ & $845_{-527}^{+762}$ & $1.60_{-0.52}^{+1.10}$ & $1.95_{-0.10}^{+0.10}$ & $3.1_{-2.8}^{+2.8}$ \\
 HD 63765 & $12.33$ & $25.5_{-9.1}^{+9.1}$ & $95_{-42}^{+82}$ & $3.20_{-1.59}^{+3.15}$ & $1.42_{-0.51}^{+0.51}$ & $1.6_{-1.5}^{+1.5}$ \\
 HD 65277A & $2.43$ & $42.9_{-5.8}^{+5.8}$ & $1368_{-501}^{+426}$ & $5.28_{-2.04}^{+2.56}$ & $1.52_{-0.20}^{+0.20}$ & $0.7_{-0.9}^{+0.9}$ \\
 HD 68978 & $8.24$ & $16.8_{-0.8}^{+0.8}$ & $63_{-20}^{+30}$ & $0.73_{-0.13}^{+0.20}$ & $1.75_{-0.08}^{+0.08}$ & $2.2_{-1.1}^{+1.1}$ \\
 HD 69830 & $7.50$ & $33.7_{-1.0}^{+1.0}$ & $234_{-106}^{+418}$ & $1.31_{-0.31}^{+0.63}$ & $1.83_{-0.06}^{+0.06}$ & $2.5_{-1.8}^{+1.8}$ \\
HD 71835 & $7.47$ & $29.7_{-0.5}^{+0.5}$ & $1123_{-595}^{+583}$ & $2.06_{-0.70}^{+1.80}$ & $1.66_{-0.03}^{+0.03}$ & $5.0_{-4.2}^{+4.2}$ \\
 HD 78429 & $13.37$ & $43.4_{-14.8}^{+14.8}$ & $19_{-8}^{+58}$ & $0.84_{-0.33}^{+5.90}$ & $3.32_{-1.13}^{+1.13}$ & $4.4_{-1.5}^{+1.5}$ \\
HD 89454 & $1.43$ & $16.4_{-1.9}^{+1.9}$ & $327_{-187}^{+278}$ & $1.07_{-0.36}^{+0.47}$ & $1.21_{-0.14}^{+0.14}$ & $0.6_{-0.5}^{+0.5}$ \\
 HD 93083 & $10.26$ & $44.6_{-6.2}^{+6.2}$ & $321_{-213}^{+448}$ & $1.75_{-0.66}^{+1.06}$ & $1.92_{-0.27}^{+0.27}$ & $2.1_{-1.9}^{+1.9}$ \\
 HD 96700 & $5.33$ & $43.6_{-2.3}^{+2.3}$ & $194_{-72}^{+177}$ & $0.79_{-0.21}^{+0.46}$ & $3.73_{-0.20}^{+0.20}$ & $6.3_{-11.6}^{+11.6}$ \\
 
    \hline
    \hline
\end{longtable}  

\newpage

\begin{longtable}{lcccccccccr}

    \caption{Basic properties of all 268 targets in the sample.}\label{tab:star_all} \\
    
    \hline 
    Simbad name & $T_{\rm{eff}}$ [K] & $T_{\rm{eff}}$ err [K] & $\rm{log} \ g$ & $\rm{log} \ g$ err & Mean $\log R_{\rm{HK}}^{\prime}$ & Mean $\log R_{\rm{HK}}^{\prime}$ err & $\rm{T_{range}}$ & $\rm{T_{range}}$ (DRS 3.5) \\
    \hline
    
BD-08 2823 & 4672.0 & 105.0 & 4.61 & 0.26 & -4.6944 & 0.0039 & 2004-01 to 2018-04 & 2004-01 to 2010-01 \\
CD-23 395 & 4673.0 & 175.0 & 4.52 & 0.47 & -4.8528 & 0.0040 & 2004-07 to 2016-10 & 2004-07 to 2013-01 \\
CD-24 10619 & 4574.0 & 155.0 & 4.68 & 0.44 & -4.5766 & 0.0036 & 2005-06 to 2021-02 & 2005-06 to 2011-07 \\
CD-26 2288 & 4924.0 & 94.0 & 4.51 & 0.28 & -4.6539 & 0.0037 & 2004-01 to 2019-10 & 2004-01 to 2011-12 \\
HD 10180 & 5911.0 & 19.0 & 4.19 & 0.03 & -4.9957 & 0.0043 & 2003-11 to 2017-08 & 2003-11 to 2015-01 \\
 HD 101930 & 5164.0 & 61.0 & 4.49 & 0.11 & -5.0048 & 0.0033 & 2004-02 to 2016-06 & 2004-02 to 2015-05 \\
HD 102117 & 5657.0 & 24.0 & 4.21 & 0.04 & -5.1215 & 0.0050 & 2004-01 to 2016-05 & 2004-01 to 2015-05 \\
 HD 102365 & 5629.0 & 29.0 & 4.35 & 0.03 & -4.9489 & 0.0030 & 2003-11 to 2017-05 & 2003-11 to 2015-05 \\
HD 103197 & 5250.0 & 60.0 & 4.51 & 0.11 & -5.0796 & 0.0059 & 2004-02 to 2018-04 & 2004-02 to 2010-04 \\
HD 103720 & 5017.0 & 88.0 & 4.58 & 0.16 & -4.4493 & 0.0038 & 2005-02 to 2021-02 & 2005-02 to 2015-05 \\
HD 103774 & 6732.0 & 56.0 & 4.29 & 0.06 & -4.7718 & 0.0073 & 2004-12 to 2018-04 & 2004-12 to 2013-02 \\
 HD 104067 & 4969.0 & 72.0 & 4.64 & 0.13 & -4.7370 & 0.0017 & 2004-02 to 2010-04 & 2004-02 to 2010-04 \\
 HD 104800 & 5697.0 & 25.0 & 4.35 & 0.02 & -4.8877 & 0.0058 & 2004-02 to 2015-01 & 2004-02 to 2015-01 \\
 HD 105690 & 5666.0 & 38.0 & 4.53 & 0.07 & -4.3063 & 0.0012 & 2009-04 to 2017-06 & 2009-04 to 2014-07 \\
 HD 106116 & 5680.0 & 15.0 & 4.28 & 0.03 & -5.0317 & 0.0041 & 2004-02 to 2017-06 & 2004-02 to 2015-05 \\
HD 10647 & 6218.0 & 20.0 & 4.30 & 0.04 & -4.7431 & 0.0024 & 2003-11 to 2020-01 & 2003-11 to 2012-01 \\
 HD 10700 & 5310.0 & 17.0 & 4.49 & 0.03 & -4.9773 & 0.0025 & 2003-09 to 2012-10 & 2003-10 to 2012-10 \\
HD 107094 & 5562.0 & 17.0 & 4.48 & 0.03 & -4.8283 & 0.0055 & 2004-01 to 2021-02 & 2004-01 to 2015-03 \\
 HD 109200 & 5056.0 & 33.0 & 4.50 & 0.08 & -4.9542 & 0.0027 & 2004-02 to 2018-05 & 2004-02 to 2015-05 \\
 HD 109271 & 5783.0 & 18.0 & 4.13 & 0.02 & -4.9963 & 0.0074 & 2005-02 to 2018-04 & 2005-02 to 2014-07 \\
HD 111232 & 5460.0 & 21.0 & 4.41 & 0.03 & -4.9832 & 0.0041 & 2004-02 to 2021-03 & 2004-02 to 2015-05 \\
 HD 111515 & 5398.0 & 18.0 & 4.47 & 0.02 & -4.9528 & 0.0043 & 2004-02 to 2017-07 & 2004-02 to 2009-03 \\
 HD 111777 & 5666.0 & 19.0 & 4.36 & 0.03 & -4.9115 & 0.0048 & 2004-02 to 2018-06 & 2004-02 to 2014-05 \\
 HD 11397 & 5564.0 & 26.0 & 4.40 & 0.04 & -4.8975 & 0.0047 & 2003-10 to 2019-09 & 2003-10 to 2014-08 \\
 HD 114076 & 5066.0 & 25.1 & 4.47 & 0.07 & -4.9911 & 0.0058 & 2004-02 to 2015-04 & 2004-02 to 2015-04 \\
 HD 114613 & 5729.0 & 17.0 & 3.84 & 0.02 & -5.1529 & 0.0041 & 2004-01 to 2019-04 & 2004-01 to 2015-05 \\
 HD 114729 & 5844.0 & 12.0 & 4.02 & 0.02 & -5.0046 & 0.0044 & 2004-02 to 2021-03 & 2004-02 to 2014-08 \\
HD 114853 & 5705.0 & 14.0 & 4.32 & 0.02 & -4.9435 & 0.0037 & 2004-01 to 2017-08 & 2004-01 to 2015-05 \\
 HD 115617 & 5559.0 & 17.0 & 4.30 & 0.03 & -5.0132 & 0.0029 & 2004-01 to 2021-03 & 2004-01 to 2015-05 \\
 HD 117207 & 5667.0 & 21.0 & 4.22 & 0.04 & -5.0881 & 0.0045 & 2004-02 to 2021-03 & 2004-02 to 2015-05 \\
HD 117618 & 5990.0 & 13.0 & 4.18 & 0.02 & -4.9790 & 0.0040 & 2004-02 to 2019-08 & 2004-02 to 2010-05 \\
 HD 119173 & 5779.0 & 44.0 & 4.11 & 0.04 & -4.8290 & 0.0050 & 2006-02 to 2020-02 & 2006-02 to 2014-03 \\
 HD 11964 & 5326.0 & 19.0 & 3.90 & 0.04 & -5.1763 & 0.0037 & 2003-10 to 2018-11 & 2003-10 to 2015-01 \\
HD 119949 & 6359.0 & 36.0 & 4.10 & 0.04 & -4.9050 & 0.0046 & 2004-02 to 2014-01 & 2004-02 to 2014-01 \\
 HD 124292 & 5443.0 & 22.0 & 4.35 & 0.04 & -5.0128 & 0.0037 & 2004-02 to 2017-08 & 2004-02 to 2015-05 \\
 HD 125072 & 4794.0 & 102.0 & 4.51 & 0.24 & -4.9238 & 0.0018 & 2004-02 to 2017-08 & 2004-02 to 2015-05 \\
 HD 125184 & 5680.0 & 30.0 & 3.99 & 0.05 & -5.0952 & 0.0039 & 2004-02 to 2017-08 & 2004-02 to 2015-05 \\
 HD 125595 & 4636.0 & 83.9 & 4.50 & 0.26 & -4.7575 & 0.0030 & 2004-05 to 2019-08 & 2004-05 to 2010-02 \\
 HD 125612A & 5913.0 & 17.0 & 4.23 & 0.03 & -4.8988 & 0.0083 & 2004-02 to 2021-03 & 2004-02 to 2015-03 \\
 HD 125881 & 6036.0 & 17.0 & 4.24 & 0.03 & -4.8726 & 0.0036 & 2004-02 to 2013-05 & 2004-02 to 2013-05 \\
 HD 126525 & 5638.0 & 13.0 & 4.28 & 0.02 & -4.9978 & 0.0045 & 2004-06 to 2019-05 & 2004-06 to 2015-05 \\
HD 126793 & 5904.0 & 33.0 & 4.23 & 0.03 & -4.8659 & 0.0048 & 2004-05 to 2015-02 & 2004-05 to 2015-02 \\
 HD 126803 & 5470.0 & 18.0 & 4.45 & 0.04 & -4.9348 & 0.0050 & 2004-02 to 2019-08 & 2004-02 to 2014-05 \\
 HD 129642 & 4919.0 & 65.0 & 4.54 & 0.16 & -5.0077 & 0.0037 & 2004-02 to 2017-08 & 2004-02 to 2015-05 \\
HD 13060 & 5255.0 & 45.0 & 4.40 & 0.09 & -4.8547 & 0.0051 & 2003-10 to 2016-09 & 2003-10 to 2015-01 \\
 HD 131653 & 5324.0 & 26.0 & 4.57 & 0.04 & -4.9976 & 0.0077 & 2004-05 to 2015-07 & 2004-05 to 2015-05 \\
HD 131664 & 5901.0 & 26.0 & 4.30 & 0.03 & -4.8432 & 0.0076 & 2004-05 to 2021-02 & 2004-05 to 2014-07 \\
HD 132569 & 4967.0 & 44.7 & 4.64 & 0.11 & -4.6443 & 0.0043 & 2004-05 to 2011-08 & 2004-05 to 2011-08 \\
 HD 133633 & 5571.0 & 19.0 & 4.41 & 0.04 & -4.9437 & 0.0053 & 2005-05 to 2018-07 & 2005-05 to 2014-05 \\
 HD 134060 & 5966.0 & 14.0 & 4.21 & 0.03 & -4.9983 & 0.0038 & 2004-02 to 2017-05 & 2004-02 to 2015-05 \\
 HD 134088 & 5675.0 & 22.0 & 4.35 & 0.03 & -4.8092 & 0.0041 & 2004-05 to 2015-06 & 2004-05 to 2015-04 \\
 HD 134606 & 5633.0 & 28.0 & 4.29 & 0.05 & -5.1077 & 0.0042 & 2004-07 to 2017-05 & 2004-07 to 2015-05 \\
 HD 134987 & 5740.0 & 23.0 & 4.17 & 0.04 & -5.1106 & 0.0043 & 2004-02 to 2016-04 & 2004-02 to 2015-05 \\
HD 135625 & 6003.0 & 14.0 & 4.08 & 0.04 & -5.0250 & 0.0087 & 2004-02 to 2021-03 & 2004-02 to 2015-05 \\
 HD 136352 & 5664.0 & 14.0 & 4.29 & 0.02 & -4.9412 & 0.0031 & 2004-05 to 2017-08 & 2004-05 to 2015-05 \\
 HD 136713 & 4994.0 & 74.0 & 4.61 & 0.14 & -4.7748 & 0.0020 & 2005-05 to 2016-09 & 2005-05 to 2015-05 \\
HD 137388 & 5240.0 & 53.0 & 4.48 & 0.11 & -4.8949 & 0.0040 & 2005-07 to 2016-08 & 2005-07 to 2014-06 \\
 HD 137676 & 5253.0 & 18.0 & 3.99 & 0.03 & -5.1132 & 0.0044 & 2004-02 to 2018-06 & 2004-02 to 2011-08 \\
HD 13808 & 5033.0 & 38.0 & 4.50 & 0.08 & -4.8921 & 0.0034 & 2003-12 to 2016-02 & 2003-12 to 2015-01 \\
 HD 1388 & 5954.0 & 10.0 & 4.19 & 0.02 & -4.9714 & 0.0038 & 2003-10 to 2017-09 & 2003-10 to 2014-12 \\
HD 141624 & 5871.0 & 30.0 & 4.22 & 0.03 & -4.9127 & 0.0050 & 2005-05 to 2018-08 & 2005-05 to 2014-07 \\
 HD 142709 & 4728.0 & 65.0 & 4.70 & 0.18 & -4.9360 & 0.0023 & 2004-07 to 2016-04 & 2004-07 to 2015-02 \\
 HD 143361 & 5503.0 & 36.0 & 4.32 & 0.06 & -5.1130 & 0.0072 & 2007-05 to 2020-03 & 2007-05 to 2014-09 \\
 HD 144628 & 5022.0 & 26.0 & 4.58 & 0.08 & -4.9259 & 0.0025 & 2004-07 to 2019-04 & 2004-07 to 2015-05 \\
HD 145377 & 6054.0 & 16.0 & 4.27 & 0.03 & -4.6128 & 0.0051 & 2005-06 to 2011-04 & 2005-06 to 2011-04 \\
 HD 145417 & 4953.0 & 48.4 & 4.98 & 0.08 & -4.8971 & 0.0032 & 2004-05 to 2014-08 & 2004-05 to 2014-08 \\
HD 14745 & 6290.0 & 39.0 & 4.37 & 0.04 & -4.8876 & 0.0073 & 2003-11 to 2021-02 & 2003-11 to 2014-03 \\
HD 148156 & 6251.0 & 25.0 & 4.18 & 0.05 & -4.9315 & 0.0087 & 2005-04 to 2019-07 & 2005-04 to 2013-07 \\
 HD 148211 & 5948.0 & 22.0 & 4.15 & 0.02 & -4.9037 & 0.0050 & 2004-05 to 2016-04 & 2004-05 to 2015-03 \\
 HD 148303 & 4829.0 & 84.0 & 4.66 & 0.22 & -4.6532 & 0.0022 & 2004-07 to 2013-05 & 2004-07 to 2013-05 \\
HD 149396 & 5657.0 & 23.0 & 4.37 & 0.03 & -4.6836 & 0.0056 & 2004-09 to 2016-05 & 2004-09 to 2013-07 \\
HD 150177 & 6216.0 & 28.0 & 3.86 & 0.03 & -4.8915 & 0.0041 & 2004-05 to 2013-07 & 2004-05 to 2013-07 \\
 HD 150433 & 5665.0 & 12.0 & 4.33 & 0.02 & -4.9512 & 0.0038 & 2005-08 to 2017-08 & 2005-08 to 2014-09 \\
 HD 15337 & 5179.0 & 44.0 & 4.48 & 0.09 & -4.9226 & 0.0039 & 2003-12 to 2019-09 & 2003-12 to 2015-01 \\
 HD 153950 & 6074.0 & 15.0 & 4.13 & 0.03 & -4.9721 & 0.0082 & 2005-05 to 2018-03 & 2005-05 to 2014-07 \\
 HD 154088 & 5374.0 & 43.0 & 4.38 & 0.07 & -5.0718 & 0.0029 & 2006-04 to 2017-08 & 2006-04 to 2015-05 \\
 HD 154577 & 4847.0 & 35.0 & 4.69 & 0.07 & -4.8699 & 0.0022 & 2004-05 to 2017-08 & 2004-05 to 2015-05 \\
HD 156098 & 6517.0 & 44.0 & 3.76 & 0.05 & -4.7816 & 0.0045 & 2005-05 to 2021-03 & 2005-05 to 2015-05 \\
 HD 15612 & 5256.0 & 34.0 & 4.55 & 0.06 & -4.5080 & 0.0044 & 2003-11 to 2018-02 & 2003-11 to 2014-03 \\
 HD 156411 & 5910.0 & 16.0 & 3.79 & 0.01 & -5.1400 & 0.0096 & 2005-05 to 2019-07 & 2005-05 to 2013-07 \\
HD 157172 & 5451.0 & 27.0 & 4.37 & 0.05 & -5.0040 & 0.0037 & 2005-07 to 2017-03 & 2005-07 to 2015-05 \\
HD 157347 & 5676.0 & 16.0 & 4.27 & 0.03 & -5.0213 & 0.0038 & 2006-03 to 2019-08 & 2006-03 to 2015-05 \\
HD 157830 & 5540.0 & 16.0 & 4.43 & 0.02 & -4.7868 & 0.0030 & 2004-05 to 2012-09 & 2004-05 to 2012-09 \\
 HD 1581 & 5951.0 & 13.0 & 4.28 & 0.03 & -4.9174 & 0.0031 & 2003-10 to 2020-12 & 2003-10 to 2014-12 \\
 HD 16008 & 5770.0 & 14.0 & 4.33 & 0.03 & -4.8465 & 0.0049 & 2003-10 to 2021-03 & 2003-10 to 2015-01 \\
 HD 161098 & 5560.0 & 15.0 & 4.40 & 0.02 & -4.9240 & 0.0037 & 2006-04 to 2017-09 & 2006-04 to 2015-05 \\
HD 161566 & 6230.0 & 24.0 & 3.88 & 0.04 & -4.9879 & 0.0077 & 2005-06 to 2019-07 & 2005-06 to 2013-03 \\
 HD 16160 & 4796.0 & 109.0 & 4.62 & 0.27 & -4.8649 & 0.0020 & 2003-10 to 2008-09 & 2003-10 to 2008-09 \\
 HD 16280 & 4677.0 & 83.5 & 4.58 & 0.31 & -4.6712 & 0.0039 & 2003-10 to 2021-03 & 2003-10 to 2014-12 \\
 HD 16417 & 5841.0 & 17.0 & 3.99 & 0.02 & -5.1028 & 0.0043 & 2003-11 to 2017-09 & 2003-11 to 2015-01 \\
 HD 165131 & 5870.0 & 15.0 & 4.27 & 0.03 & -4.9658 & 0.0086 & 2006-05 to 2021-03 & 2006-05 to 2015-05 \\
HD 166724 & 5099.0 & 36.0 & 4.55 & 0.09 & -4.7253 & 0.0033 & 2004-05 to 2011-09 & 2004-05 to 2011-09 \\
 HD 167677 & 5474.0 & 20.0 & 4.40 & 0.03 & -4.9899 & 0.0074 & 2005-05 to 2019-08 & 2005-05 to 2015-05 \\
 HD 168863 & 4782.0 & 80.0 & 4.52 & 0.23 & -4.7762 & 0.0041 & 2006-07 to 2021-03 & 2006-07 to 2014-07 \\
 HD 170493 & 4751.0 & 108.0 & 4.49 & 0.25 & -4.8776 & 0.0024 & 2005-07 to 2015-09 & 2005-07 to 2015-05 \\
HD 171028 & 5671.0 & 16.0 & 3.73 & 0.03 & -5.1098 & 0.0071 & 2004-10 to 2017-09 & 2004-10 to 2011-03 \\
 HD 171587 & 5412.0 & 15.0 & 4.58 & 0.02 & -4.7553 & 0.0035 & 2004-11 to 2014-08 & 2004-11 to 2014-08 \\
HD 172513 & 5500.0 & 18.0 & 4.37 & 0.03 & -4.7730 & 0.0025 & 2004-05 to 2007-05 & 2004-05 to 2007-05 \\
 HD 172568 & 5728.0 & 22.0 & 4.45 & 0.03 & -4.9046 & 0.0051 & 2005-05 to 2018-08 & 2005-05 to 2013-10 \\
 HD 175607 & 5392.0 & 17.0 & 4.51 & 0.03 & -4.9298 & 0.0052 & 2004-07 to 2015-04 & 2004-07 to 2015-04 \\
HD 176354 & 5271.0 & 40.0 & 3.89 & 0.07 & -5.0505 & 0.0052 & 2006-05 to 2019-07 & 2006-05 to 2011-09 \\
 HD 176986 & 5018.0 & 59.0 & 4.60 & 0.11 & -4.8283 & 0.0028 & 2004-07 to 2017-08 & 2004-07 to 2015-04 \\
 HD 177565 & 5627.0 & 19.0 & 4.30 & 0.03 & -4.9085 & 0.0027 & 2003-10 to 2017-09 & 2003-10 to 2015-05 \\
 HD 17865 & 5877.0 & 24.0 & 4.13 & 0.03 & -4.9272 & 0.0051 & 2003-11 to 2014-09 & 2003-11 to 2014-09 \\
 HD 17970 & 5038.0 & 31.0 & 4.53 & 0.06 & -5.0043 & 0.0037 & 2003-10 to 2017-09 & 2003-10 to 2015-01 \\
 HD 181433 & 4962.0 & 134.0 & 4.54 & 0.27 & -5.1290 & 0.0036 & 2003-10 to 2017-08 & 2003-10 to 2015-05 \\
 HD 181720 & 5792.0 & 17.0 & 4.10 & 0.02 & -4.9792 & 0.0051 & 2005-05 to 2017-11 & 2005-05 to 2014-08 \\
 HD 183658 & 5803.0 & 17.0 & 4.24 & 0.02 & -4.9407 & 0.0043 & 2005-08 to 2019-05 & 2005-08 to 2015-05 \\
HD 185283 & 4746.0 & 79.3 & 4.53 & 0.20 & -4.9514 & 0.0046 & 2005-07 to 2019-09 & 2005-07 to 2015-05 \\
 HD 189567 & 5726.0 & 15.0 & 4.28 & 0.01 & -4.9120 & 0.0031 & 2003-10 to 2017-08 & 2003-10 to 2015-05 \\
 HD 190248 & 5644.0 & 30.0 & 4.12 & 0.05 & -5.1315 & 0.0034 & 2003-10 to 2016-10 & 2003-10 to 2015-05 \\
 HD 190647 & 5639.0 & 24.0 & 4.09 & 0.04 & -5.1613 & 0.0064 & 2004-09 to 2021-03 & 2004-09 to 2014-10 \\
HD 190984 & 6007.0 & 25.0 & 3.78 & 0.03 & -5.0626 & 0.0073 & 2004-05 to 2013-08 & 2004-05 to 2013-08 \\
HD 191797 & 5037.0 & 48.5 & 4.61 & 0.11 & -4.3751 & 0.0037 & 2004-07 to 2019-08 & 2004-07 to 2014-07 \\
 HD 192310 & 5087.0 & 48.0 & 4.42 & 0.10 & -4.9509 & 0.0019 & 2003-11 to 2016-10 & 2003-11 to 2014-10 \\
 HD 19467 & 5720.0 & 10.0 & 4.18 & 0.01 & -5.0079 & 0.0040 & 2003-10 to 2017-10 & 2003-10 to 2015-01 \\
 HD 195145 & 5625.0 & 20.0 & 4.32 & 0.03 & -4.9820 & 0.0081 & 2006-07 to 2018-06 & 2006-07 to 2013-07 \\
HD 19641 & 5806.0 & 14.0 & 4.23 & 0.02 & -4.9734 & 0.0082 & 2003-11 to 2021-01 & 2003-11 to 2014-12 \\
 HD 197027 & 5694.0 & 28.0 & 4.09 & 0.04 & -5.0001 & 0.0067 & 2009-05 to 2019-05 & 2009-05 to 2014-11 \\
 HD 197197 & 5812.0 & 16.0 & 4.04 & 0.02 & -5.0203 & 0.0061 & 2003-11 to 2018-10 & 2003-11 to 2015-04 \\
 HD 199288 & 5765.0 & 19.0 & 4.36 & 0.02 & -4.8868 & 0.0033 & 2003-11 to 2019-11 & 2003-11 to 2015-05 \\
 HD 199289 & 5928.0 & 37.0 & 4.43 & 0.03 & -4.8466 & 0.0050 & 2003-11 to 2018-08 & 2003-11 to 2014-10 \\
 HD 199604 & 5817.0 & 22.0 & 4.18 & 0.03 & -4.9118 & 0.0057 & 2003-11 to 2018-10 & 2003-11 to 2010-11 \\
HD 199847 & 5763.0 & 20.0 & 4.08 & 0.02 & -5.0033 & 0.0067 & 2003-11 to 2018-10 & 2003-11 to 2014-09 \\
HD 20003 & 5494.0 & 27.0 & 4.37 & 0.05 & -4.9891 & 0.0050 & 2003-12 to 2017-08 & 2003-12 to 2015-01 \\
HD 200538 & 6042.0 & 18.0 & 4.13 & 0.03 & -5.0017 & 0.0085 & 2004-09 to 2017-04 & 2004-09 to 2012-09 \\
HD 202206 & 5757.0 & 25.0 & 4.33 & 0.03 & -4.7434 & 0.0028 & 2004-05 to 2019-06 & 2004-05 to 2012-06 \\
HD 202871 & 6055.0 & 16.0 & 4.28 & 0.04 & -4.7825 & 0.0075 & 2005-06 to 2012-09 & 2005-06 to 2012-09 \\
 HD 20407 & 5866.0 & 14.0 & 4.32 & 0.01 & -4.8786 & 0.0034 & 2003-10 to 2017-11 & 2003-10 to 2015-02 \\
 HD 204313 & 5776.0 & 22.0 & 4.23 & 0.02 & -5.0384 & 0.0046 & 2006-05 to 2017-09 & 2006-05 to 2014-10 \\
 HD 204941 & 4997.0 & 36.0 & 4.52 & 0.10 & -4.9275 & 0.0034 & 2004-12 to 2016-09 & 2004-12 to 2014-10 \\
HD 205294 & 6370.0 & 32.0 & 3.92 & 0.04 & -4.8884 & 0.0072 & 2003-10 to 2018-09 & 2003-10 to 2011-08 \\
 HD 205536 & 5442.0 & 23.0 & 4.36 & 0.04 & -5.0377 & 0.0036 & 2003-10 to 2020-11 & 2003-10 to 2009-11 \\
 HD 206998 & 5822.0 & 26.0 & 4.08 & 0.03 & -4.9271 & 0.0058 & 2003-11 to 2018-10 & 2003-11 to 2014-10 \\
 HD 2071 & 5719.0 & 14.0 & 4.35 & 0.02 & -4.9186 & 0.0036 & 2003-10 to 2017-10 & 2003-10 to 2014-11 \\
 HD 207129 & 5937.0 & 13.0 & 4.28 & 0.02 & -4.8853 & 0.0030 & 2003-10 to 2020-11 & 2003-10 to 2015-05 \\
 HD 20781 & 5256.0 & 29.0 & 4.43 & 0.05 & -5.0498 & 0.0043 & 2003-12 to 2017-09 & 2003-12 to 2015-02 \\
 HD 20782 & 5774.0 & 14.0 & 4.22 & 0.01 & -4.9044 & 0.0036 & 2003-10 to 2017-10 & 2003-10 to 2015-01 \\
 HD 207832 & 5718.0 & 27.0 & 4.33 & 0.04 & -4.7324 & 0.0052 & 2003-10 to 2019-09 & 2003-10 to 2013-10 \\
 HD 207869 & 5527.0 & 21.0 & 4.45 & 0.05 & -4.9486 & 0.0060 & 2003-11 to 2017-08 & 2003-11 to 2013-08 \\
 HD 20807 & 5824.0 & 15.0 & 4.29 & 0.03 & -4.8720 & 0.0029 & 2003-10 to 2021-02 & 2003-10 to 2015-02 \\
 HD 208487 & 6146.0 & 19.0 & 4.19 & 0.03 & -4.9230 & 0.0038 & 2004-09 to 2019-09 & 2004-09 to 2007-06 \\
 HD 20868 & 4720.0 & 91.0 & 4.50 & 0.22 & -4.9930 & 0.0050 & 2003-11 to 2019-08 & 2003-11 to 2008-08 \\
HD 208704 & 5826.0 & 11.0 & 4.21 & 0.01 & -4.9438 & 0.0037 & 2004-07 to 2017-10 & 2004-07 to 2014-11 \\
 HD 210918 & 5755.0 & 12.0 & 4.21 & 0.02 & -5.0101 & 0.0037 & 2003-10 to 2019-12 & 2003-10 to 2014-09 \\
 HD 211038 & 4974.0 & 17.0 & 3.84 & 0.05 & -5.1615 & 0.0030 & 2003-11 to 2017-08 & 2003-11 to 2014-10 \\
HD 21132 & 6243.0 & 34.0 & 4.27 & 0.05 & -4.8531 & 0.0058 & 2004-02 to 2020-01 & 2004-02 to 2015-01 \\
 HD 21209 & 4671.0 & 65.0 & 4.59 & 0.15 & -4.7945 & 0.0026 & 2003-10 to 2016-09 & 2003-10 to 2015-01 \\
 HD 215152 & 4803.0 & 52.0 & 4.49 & 0.15 & -4.8599 & 0.0022 & 2003-12 to 2016-09 & 2003-12 to 2014-08 \\
 HD 215456 & 5789.0 & 15.0 & 3.95 & 0.03 & -5.1311 & 0.0049 & 2003-10 to 2017-05 & 2003-10 to 2014-10 \\
HD 215497 & 5003.0 & 103.0 & 4.41 & 0.26 & -5.0695 & 0.0050 & 2004-10 to 2018-08 & 2004-10 to 2009-11 \\
 HD 216770 & 5424.0 & 51.0 & 4.37 & 0.07 & -4.9645 & 0.0031 & 2003-10 to 2019-09 & 2003-10 to 2013-11 \\
 HD 21693 & 5430.0 & 26.0 & 4.36 & 0.04 & -4.9244 & 0.0035 & 2003-10 to 2017-09 & 2003-10 to 2015-01 \\
 HD 21749 & 4640.0 & 100.0 & 4.91 & 0.07 & -4.6899 & 0.0019 & 2003-11 to 2016-12 & 2003-11 to 2009-12 \\
HD 218504 & 5962.0 & 29.0 & 4.12 & 0.03 & -4.9327 & 0.0045 & 2003-11 to 2014-09 & 2003-11 to 2014-09 \\
HD 219828 & 5888.0 & 14.0 & 4.01 & 0.02 & -5.1190 & 0.0064 & 2005-05 to 2015-08 & 2005-05 to 2013-11 \\
 HD 220339 & 4938.0 & 32.0 & 4.60 & 0.08 & -4.8213 & 0.0024 & 2003-10 to 2020-12 & 2003-10 to 2014-12 \\
 HD 22049 & 5049.0 & 48.0 & 4.59 & 0.09 & -4.4937 & 0.0007 & 2003-11 to 2020-01 & 2003-11 to 2007-08 \\
HD 220507 & 5698.0 & 17.0 & 4.17 & 0.03 & -5.0747 & 0.0049 & 2003-10 to 2017-10 & 2003-10 to 2014-12 \\
HD 221580 & 5322.0 & 24.0 & 2.71 & 0.04 & -5.4860 & 0.0158 & 2003-11 to 2007-10 & 2003-11 to 2007-10 \\
HD 222669 & 5894.0 & 17.0 & 4.27 & 0.02 & -4.8376 & 0.0038 & 2003-12 to 2016-10 & 2003-12 to 2014-12 \\
 HD 223171 & 5841.0 & 18.0 & 4.03 & 0.02 & -5.0245 & 0.0042 & 2003-10 to 2018-11 & 2003-10 to 2014-12 \\
 HD 224685 & 5504.0 & 30.0 & 4.43 & 0.06 & -4.9248 & 0.0064 & 2004-12 to 2018-08 & 2004-12 to 2014-12 \\
 HD 224817 & 5894.0 & 22.0 & 4.17 & 0.02 & -4.9232 & 0.0052 & 2003-11 to 2015-07 & 2003-11 to 2014-11 \\
HD 22879 & 5857.0 & 27.0 & 4.28 & 0.02 & -4.8791 & 0.0035 & 2003-10 to 2017-10 & 2003-10 to 2015-01 \\
 HD 23249 & 5035.0 & 39.0 & 3.80 & 0.08 & -5.1648 & 0.0024 & 2003-10 to 2017-03 & 2003-10 to 2015-01 \\
HD 24062 & 6107.0 & 60.0 & 4.34 & 0.06 & -5.0123 & 0.0086 & 2003-11 to 2021-03 & 2003-11 to 2013-11 \\
 HD 24633 & 5276.0 & 43.0 & 4.41 & 0.07 & -4.9850 & 0.0071 & 2003-11 to 2021-03 & 2003-11 to 2013-12 \\
HD 25171 & 6160.0 & 24.0 & 4.13 & 0.03 & -4.9322 & 0.0068 & 2003-11 to 2020-12 & 2003-11 to 2015-02 \\
 HD 26965 & 5072.0 & 53.0 & 4.58 & 0.19 & -4.9495 & 0.0022 & 2003-10 to 2021-02 & 2003-10 to 2015-01 \\
HD 27063 & 5767.0 & 14.0 & 4.30 & 0.03 & -4.7496 & 0.0029 & 2003-11 to 2008-02 & 2003-11 to 2008-02 \\
 HD 27894 & 4952.0 & 105.0 & 4.56 & 0.20 & -4.9392 & 0.0065 & 2003-10 to 2019-03 & 2003-10 to 2013-11 \\
 HD 28254A & 5653.0 & 33.0 & 4.05 & 0.05 & -5.1685 & 0.0094 & 2003-10 to 2021-03 & 2003-10 to 2014-09 \\
 HD 28471 & 5745.0 & 14.0 & 4.24 & 0.01 & -4.9963 & 0.0043 & 2003-11 to 2018-01 & 2003-11 to 2013-01 \\
 HD 297396 & 4622.0 & 114.0 & 4.58 & 0.36 & -4.8059 & 0.0039 & 2004-01 to 2021-02 & 2004-01 to 2015-05 \\
HD 31103 & 6078.0 & 16.0 & 4.23 & 0.02 & -4.6294 & 0.0055 & 2003-11 to 2020-11 & 2003-11 to 2013-11 \\
 HD 31128 & 6096.0 & 67.0 & 4.63 & 0.06 & -4.8146 & 0.0057 & 2003-11 to 2015-04 & 2003-11 to 2015-04 \\
HD 31527 & 5898.0 & 13.0 & 4.26 & 0.02 & -4.9395 & 0.0038 & 2003-10 to 2017-08 & 2003-10 to 2015-02 \\
HD 31822 & 6042.0 & 16.0 & 4.32 & 0.03 & -4.8204 & 0.0035 & 2003-10 to 2016-03 & 2003-10 to 2015-02 \\
HD 3220 & 5846.0 & 15.0 & 4.34 & 0.02 & -4.8183 & 0.0079 & 2003-11 to 2010-11 & 2003-11 to 2010-11 \\
HD 32564 & 5533.0 & 29.0 & 4.31 & 0.06 & -5.0305 & 0.0052 & 2009-11 to 2016-03 & 2009-11 to 2015-01 \\
 HD 330075 & 4958.0 & 52.0 & 4.41 & 0.13 & -4.9777 & 0.0041 & 2004-02 to 2015-04 & 2004-02 to 2012-02 \\
 HD 35854 & 4928.0 & 56.0 & 4.64 & 0.11 & -4.8052 & 0.0021 & 2003-10 to 2018-05 & 2003-10 to 2015-02 \\
 HD 36003 & 4647.0 & 88.0 & 4.60 & 0.21 & -4.8609 & 0.0016 & 2003-12 to 2016-03 & 2003-12 to 2015-03 \\
 HD 36379 & 6030.0 & 14.0 & 4.05 & 0.02 & -4.9482 & 0.0040 & 2003-10 to 2017-04 & 2003-10 to 2015-03 \\
 HD 3823 & 6022.0 & 14.0 & 4.07 & 0.02 & -4.9692 & 0.0039 & 2003-10 to 2017-09 & 2003-10 to 2015-01 \\
 HD 38858 & 5733.0 & 12.0 & 4.38 & 0.01 & -4.9051 & 0.0029 & 2003-10 to 2019-01 & 2003-10 to 2015-02 \\
 HD 39091 & 6003.0 & 17.0 & 4.18 & 0.03 & -4.9815 & 0.0037 & 2003-12 to 2020-12 & 2003-12 to 2015-01 \\
 HD 39194 & 5205.0 & 23.0 & 4.61 & 0.05 & -4.9628 & 0.0041 & 2003-11 to 2017-08 & 2003-11 to 2015-02 \\
 HD 3964 & 5729.0 & 19.0 & 4.37 & 0.04 & -4.8328 & 0.0073 & 2003-11 to 2020-12 & 2003-11 to 2014-09 \\
HD 40307 & 4977.0 & 59.0 & 4.63 & 0.16 & -4.9389 & 0.0021 & 2003-10 to 2019-04 & 2003-10 to 2015-02 \\
 HD 40397 & 5527.0 & 20.0 & 4.34 & 0.04 & -5.0316 & 0.0037 & 2003-10 to 2016-09 & 2003-10 to 2015-04 \\
 HD 40865 & 5719.0 & 16.0 & 4.38 & 0.03 & -4.9144 & 0.0053 & 2003-10 to 2017-12 & 2003-10 to 2014-12 \\
HD 41248 & 5713.0 & 21.0 & 4.37 & 0.03 & -4.8806 & 0.0053 & 2003-10 to 2017-12 & 2003-10 to 2014-01 \\
 HD 4308 & 5644.0 & 16.0 & 4.28 & 0.03 & -4.9553 & 0.0035 & 2003-10 to 2021-01 & 2003-10 to 2015-01 \\
HD 43197 & 5449.0 & 42.0 & 4.28 & 0.08 & -5.0857 & 0.0070 & 2003-12 to 2021-03 & 2003-12 to 2013-03 \\
HD 44219 & 5766.0 & 18.0 & 4.06 & 0.03 & -5.0551 & 0.0083 & 2003-11 to 2021-03 & 2003-11 to 2013-03 \\
 HD 45184 & 5869.0 & 14.0 & 4.29 & 0.02 & -4.9011 & 0.0031 & 2003-10 to 2021-03 & 2003-10 to 2015-03 \\
HD 45364 & 5434.0 & 20.0 & 4.37 & 0.03 & -4.9753 & 0.0041 & 2003-12 to 2017-09 & 2003-12 to 2015-02 \\
HD 457 & 6089.0 & 23.0 & 4.16 & 0.03 & -5.0230 & 0.0085 & 2003-10 to 2020-11 & 2003-10 to 2014-12 \\
 HD 47186 & 5675.0 & 21.0 & 4.25 & 0.04 & -5.0712 & 0.0043 & 2003-12 to 2017-09 & 2003-12 to 2015-04 \\
HD 48115 & 5825.0 & 12.0 & 4.31 & 0.02 & -4.7526 & 0.0070 & 2004-01 to 2020-11 & 2004-01 to 2011-01 \\
HD 48265 & 5798.0 & 29.0 & 3.79 & 0.14 & -5.1912 & 0.0063 & 2007-09 to 2021-03 & 2007-09 to 2015-02 \\
 HD 4915 & 5658.0 & 13.0 & 4.42 & 0.03 & -4.7879 & 0.0025 & 2003-10 to 2020-12 & 2003-10 to 2008-09 \\
 HD 51608 & 5358.0 & 22.0 & 4.38 & 0.05 & -5.0082 & 0.0043 & 2003-12 to 2017-09 & 2003-12 to 2015-04 \\
HD 52265 & 6136.0 & 31.0 & 4.07 & 0.03 & -4.9915 & 0.0043 & 2004-02 to 2021-02 & 2004-02 to 2010-04 \\
 HD 5388 & 6311.0 & 33.0 & 3.89 & 0.03 & -4.9165 & 0.0082 & 2003-11 to 2019-07 & 2003-11 to 2010-01 \\
 HD 56274 & 5734.0 & 22.0 & 4.38 & 0.03 & -4.8320 & 0.0034 & 2003-11 to 2015-04 & 2003-11 to 2015-04 \\
 HD 59468 & 5618.0 & 20.0 & 4.30 & 0.03 & -5.0095 & 0.0034 & 2003-10 to 2018-05 & 2003-10 to 2015-05 \\
 HD 59711A & 5722.0 & 13.0 & 4.33 & 0.02 & -4.9295 & 0.0041 & 2003-10 to 2017-11 & 2003-10 to 2015-01 \\
HD 60532 & 6273.0 & 37.0 & 3.68 & 0.04 & -5.0976 & 0.0050 & 2006-02 to 2011-06 & 2006-02 to 2011-06 \\
 HD 61051 & 5363.0 & 27.0 & 4.38 & 0.05 & -5.0692 & 0.0079 & 2004-01 to 2021-01 & 2004-01 to 2015-03 \\
 HD 61383 & 5716.0 & 14.0 & 4.08 & 0.02 & -5.0179 & 0.0088 & 2003-12 to 2021-03 & 2003-12 to 2015-03 \\
 HD 61986 & 5725.0 & 20.0 & 4.35 & 0.04 & -4.8898 & 0.0045 & 2003-12 to 2017-05 & 2003-12 to 2010-12 \\
 HD 63765 & 5432.0 & 19.0 & 4.41 & 0.03 & -4.7358 & 0.0026 & 2003-12 to 2019-04 & 2003-12 to 2010-04 \\
 HD 65277A & 4701.0 & 57.0 & 4.57 & 0.16 & -5.0334 & 0.0025 & 2003-12 to 2017-01 & 2003-12 to 2015-05 \\
 HD 65907A & 5945.0 & 16.0 & 4.31 & 0.02 & -4.9009 & 0.0032 & 2003-10 to 2021-01 & 2003-10 to 2015-04 \\
 HD 68146 & 6427.0 & 44.0 & 4.12 & 0.04 & -4.8606 & 0.0033 & 2006-02 to 2014-05 & 2006-02 to 2014-05 \\
HD 68284 & 5933.0 & 26.0 & 3.87 & 0.03 & -5.0952 & 0.0062 & 2003-11 to 2015-01 & 2003-11 to 2015-01 \\
HD 68607 & 5215.0 & 45.0 & 4.48 & 0.08 & -4.7254 & 0.0027 & 2003-12 to 2007-04 & 2003-12 to 2007-04 \\
 HD 68978 & 5965.0 & 22.0 & 4.26 & 0.02 & -4.8548 & 0.0029 & 2003-10 to 2016-03 & 2003-10 to 2015-05 \\
 HD 69611 & 5762.0 & 25.0 & 4.17 & 0.03 & -4.9562 & 0.0045 & 2003-11 to 2017-04 & 2003-11 to 2015-01 \\
 HD 69830 & 5402.0 & 28.0 & 4.40 & 0.04 & -4.9989 & 0.0027 & 2003-10 to 2019-12 & 2003-10 to 2015-05 \\
 HD 70642 & 5668.0 & 22.0 & 4.30 & 0.04 & -5.0217 & 0.0046 & 2003-12 to 2021-03 & 2003-12 to 2015-03 \\
 HD 71334 & 5694.0 & 13.0 & 4.26 & 0.03 & -4.9853 & 0.0045 & 2003-12 to 2017-02 & 2003-12 to 2015-01 \\
HD 71835 & 5438.0 & 22.0 & 4.37 & 0.04 & -4.9417 & 0.0037 & 2003-12 to 2017-05 & 2003-12 to 2015-05 \\
HD 7199 & 5386.0 & 45.0 & 4.34 & 0.08 & -4.9891 & 0.0037 & 2003-11 to 2017-09 & 2003-11 to 2015-01 \\
HD 72659 & 5926.0 & 12.0 & 4.03 & 0.01 & -4.9963 & 0.0086 & 2004-02 to 2021-03 & 2004-02 to 2014-06 \\
 HD 72673 & 5243.0 & 22.0 & 4.52 & 0.04 & -4.9259 & 0.0022 & 2003-12 to 2019-04 & 2003-12 to 2015-05 \\
 HD 73267 & 5373.0 & 30.0 & 4.38 & 0.05 & -5.0939 & 0.0067 & 2004-01 to 2021-03 & 2004-01 to 2015-05 \\
 HD 73524 & 6017.0 & 13.0 & 4.19 & 0.03 & -4.9962 & 0.0040 & 2003-12 to 2019-04 & 2003-12 to 2015-05 \\
HD 73583 & 4597.0 & 69.0 & 4.57 & 0.28 & -4.4603 & 0.0032 & 2004-02 to 2020-03 & 2004-02 to 2005-12 \\
 HD 7449 & 6024.0 & 13.0 & 4.27 & 0.03 & -4.8206 & 0.0032 & 2003-11 to 2016-01 & 2003-11 to 2015-01 \\
 HD 74698 & 5783.0 & 19.0 & 4.12 & 0.02 & -5.0331 & 0.0090 & 2004-02 to 2021-03 & 2004-02 to 2015-05 \\
HD 74957 & 5915.0 & 20.0 & 4.34 & 0.03 & -4.9276 & 0.0081 & 2004-01 to 2021-03 & 2004-01 to 2015-02 \\
 HD 76151 & 5788.0 & 23.0 & 4.33 & 0.02 & -4.7266 & 0.0022 & 2003-11 to 2021-02 & 2003-11 to 2012-11 \\
 HD 77110 & 5717.0 & 20.0 & 4.36 & 0.02 & -4.9256 & 0.0050 & 2004-01 to 2017-05 & 2004-01 to 2014-02 \\
 HD 77338 & 5440.0 & 52.0 & 4.34 & 0.11 & -5.0587 & 0.0047 & 2004-12 to 2019-06 & 2004-12 to 2013-09 \\
 HD 78429 & 5760.0 & 19.0 & 4.19 & 0.02 & -4.9518 & 0.0037 & 2003-12 to 2017-06 & 2003-12 to 2015-05 \\
HD 79601 & 5825.0 & 25.0 & 4.15 & 0.03 & -4.9291 & 0.0047 & 2004-01 to 2015-02 & 2004-01 to 2015-02 \\
 HD 82342 & 4470.0 & 21.0 & 4.96 & 0.05 & -4.9746 & 0.0030 & 2003-12 to 2016-05 & 2003-12 to 2015-05 \\
 HD 82516 & 5041.0 & 57.0 & 4.61 & 0.12 & -4.9816 & 0.0031 & 2003-12 to 2017-07 & 2003-12 to 2015-05 \\
 HD 82943 & 5989.0 & 20.0 & 4.20 & 0.02 & -4.9623 & 0.0034 & 2004-01 to 2017-06 & 2004-01 to 2015-05 \\
 HD 8389A & 5283.0 & 64.0 & 4.41 & 0.12 & -5.0311 & 0.0028 & 2003-11 to 2020-12 & 2003-11 to 2013-09 \\
HD 8535 & 6158.0 & 13.0 & 4.12 & 0.02 & -4.9292 & 0.0081 & 2003-11 to 2019-08 & 2003-11 to 2014-10 \\
HD 85390 & 5135.0 & 45.0 & 4.42 & 0.12 & -4.9666 & 0.0037 & 2003-12 to 2017-05 & 2003-12 to 2015-05 \\
 HD 85725 & 5986.0 & 26.0 & 3.72 & 0.04 & -5.2147 & 0.0093 & 2004-02 to 2021-02 & 2004-02 to 2015-05 \\
 HD 87838 & 6118.0 & 33.0 & 4.19 & 0.03 & -4.8489 & 0.0045 & 2004-01 to 2017-07 & 2004-01 to 2015-05 \\
 HD 88218 & 5878.0 & 14.0 & 3.97 & 0.02 & -5.0915 & 0.0045 & 2003-12 to 2016-05 & 2003-12 to 2015-05 \\
 HD 8828 & 5403.0 & 25.0 & 4.46 & 0.03 & -5.0145 & 0.0042 & 2003-10 to 2017-09 & 2003-10 to 2015-01 \\
 HD 88725 & 5654.0 & 17.0 & 4.39 & 0.03 & -4.8848 & 0.0039 & 2004-02 to 2015-06 & 2004-02 to 2014-12 \\
HD 89454 & 5728.0 & 17.0 & 4.34 & 0.03 & -4.6977 & 0.0024 & 2004-01 to 2008-01 & 2004-01 to 2008-01 \\
HD 89839 & 6314.0 & 24.0 & 4.13 & 0.06 & -4.9294 & 0.0080 & 2004-02 to 2021-02 & 2004-02 to 2015-05 \\
 HD 90156 & 5599.0 & 12.0 & 4.40 & 0.02 & -4.9509 & 0.0034 & 2004-01 to 2017-07 & 2004-01 to 2015-05 \\
 HD 91889 & 6140.0 & 22.0 & 4.03 & 0.03 & -4.8652 & 0.0031 & 2006-02 to 2011-02 & 2006-02 to 2011-02 \\
 HD 92719 & 5824.0 & 16.0 & 4.34 & 0.03 & -4.8249 & 0.0027 & 2004-01 to 2016-05 & 2004-01 to 2015-04 \\
 HD 93083 & 5048.0 & 66.0 & 4.46 & 0.16 & -4.9899 & 0.0031 & 2004-01 to 2017-07 & 2004-01 to 2015-04 \\
HD 93385 & 5977.0 & 18.0 & 4.19 & 0.02 & -4.9714 & 0.0043 & 2003-12 to 2017-05 & 2003-12 to 2015-04 \\
 HD 94151 & 5583.0 & 19.0 & 4.31 & 0.02 & -4.9724 & 0.0038 & 2003-12 to 2019-04 & 2003-12 to 2013-03 \\
HD 94771 & 5631.0 & 21.0 & 3.94 & 0.03 & -5.2218 & 0.0090 & 2004-02 to 2021-03 & 2004-02 to 2015-05 \\
HD 95456 & 6276.0 & 22.0 & 4.01 & 0.04 & -4.9432 & 0.0038 & 2003-12 to 2017-06 & 2003-12 to 2015-05 \\
HD 95542 & 5984.0 & 15.0 & 4.29 & 0.03 & -4.6653 & 0.0068 & 2004-01 to 2020-12 & 2004-01 to 2015-03 \\
HD 9578 & 6055.0 & 14.0 & 4.26 & 0.03 & -4.5938 & 0.0055 & 2003-11 to 2020-11 & 2003-11 to 2012-11 \\
HD 96423 & 5711.0 & 18.0 & 4.23 & 0.02 & -5.0486 & 0.0042 & 2003-12 to 2019-05 & 2003-12 to 2015-01 \\
 HD 96700 & 5845.0 & 13.0 & 4.22 & 0.02 & -4.9378 & 0.0035 & 2004-01 to 2017-07 & 2004-01 to 2015-04 \\
 HD 97037 & 5883.0 & 14.0 & 4.15 & 0.02 & -4.9815 & 0.0039 & 2004-01 to 2017-02 & 2004-01 to 2015-01 \\
 HD 97343 & 5410.0 & 20.0 & 4.39 & 0.03 & -5.0374 & 0.0033 & 2004-01 to 2018-05 & 2004-01 to 2015-05 \\
 HD 98281 & 5381.0 & 23.0 & 4.43 & 0.04 & -4.9154 & 0.0029 & 2004-01 to 2016-03 & 2004-01 to 2015-03 \\

    \hline
    \hline
\end{longtable}

\bsp	
\label{lastpage}
\end{document}